\documentclass{aa}  

\usepackage{graphicx}
\usepackage{txfonts}
\usepackage{hyperref}
\newcommand{\Mearth}{M_{\oplus}}
\newcommand{\Msun}{M_{\odot}}
\newcommand{\Myr}{\mathrm{\, Myr}}

\newcommand{\CO}{\ensuremath{\mathrm{CO}}}
\newcommand{\CN}{\ensuremath{\mathrm{CN}}}
\newcommand{\HCOp}{\ensuremath{\mathrm{HCO}^{+}}}
\newcommand{\CI}{\ensuremath{\ion{C}{i}}}
\newcommand{\tCO}{\ensuremath{{}^{13}\mathrm{CO}}}
\newcommand{\Htp}{\ensuremath{\mathrm{H}_3^{+}}}
\newcommand{\CHtp}{\ensuremath{\mathrm{CH}_3^{+}}}
\begin{document}

   \title{Thermochemical constraints on a primordial-origin of gas-rich debris disks}

   \author{H. Mitani\inst{1,2}\corrauth{hiroto.mitani@astr.tohoku.ac.jp}
   \fnmsep\thanks{\emph{Present address:} Astronomical Institute, Graduate School of Science, Tohoku University, 6-3 Aoba, Aramaki, Aoba-ku, Sendai, Miyagi 980-8578, Japan} 
   \and W. Ooyama\inst{3} 
   \and R. Nakatani\inst{4} 
   \and R. Kuiper\inst{1} 
   \and T. Hosokawa\inst{3}
   }
   
\institute{
Faculty of Physics, University of Duisburg-Essen, Lotharstra{\ss}e 1, D-47057 Duisburg, Germany
\and
Department of Physics, School of Science, The University of Tokyo, 7-3-1 Hongo, Bunkyo, Tokyo 113-0033, Japan
\and
Department of Physics, Graduate School of Science, Kyoto University, Sakyo, Kyoto 606-8502, Japan
\and
Dipartimento di Fisica, Universit\`a degli Studi di Milano, Via Celoria 16, I-20133 Milano, Italy
}
   \date{}
\titlerunning{Modelling of gas-rich debris disks}
\authorrunning{Mitani et al.}
% \abstract{}{}{}{}{} 
% 5 {} token are mandatory
 
  \abstract{Recent observations have revealed gas-rich debris disks around intermediate-mass stars at ages of tens of Myr. The origin of this gas remains unclear: it may be primordial, retained from the protoplanetary phase, or secondary, released from volatile-rich solids. Secondary-origin models reproduce CO emission but often overpredict neutral carbon. Recent observations and disk-evolution models suggest that primordial gas may survive longer than previously assumed, motivating thermochemical tests of the primordial-remnant scenario for gas-rich debris disks.} 
  {We test the previously unexplored possibility that primordial-origin disks satisfy the observational constraints on gas-rich debris disks. Specifically, we determine under what conditions a disk around a $\sim2\,\Msun$ star reproduces substantial CO, low \CI/\CO{} ratios, and weak \HCOp{} emission consistent with current non-detections.}
  {We post-processed 20--40\,Myr structures from 1D disk-evolution models with \textsc{Cloudy}, varying irradiation geometry, dust-to-gas mass ratio (DTG), and cosmic-ray ionisation rate, and calculated radial intensity profiles and disk-integrated masses.}
  {In the dust-poor externally irradiated models (${\rm DTG}\sim10^{-4}$), CO remains optically thick around $R\sim100$~au. The models yield low disk-integrated \CI/\CO{} mass ratios, $M_{\CI}/M_{\CO}\lesssim0.1$. Our model produces CO radial intensities of the observed order of magnitude, but its \CI-emitting region extends beyond $\sim300$~au and is more extended than observed. The standard CR model overproduces \HCOp, whereas $\zeta_{\rm CR}\sim10^{-19}\,\mathrm{s}^{-1}$ brings its predicted $J=2\rightarrow1$ luminosity within current observational limits.}
  {These results demonstrate that a primordial-remnant origin remains chemically viable for CO-rich debris disks. The main remaining tension is the excessive radial extent of the \CI{} emission, although it may reflect our simplified radiative transfer treatment. Further testing of the primordial-origin scenario will require multidimensional, self-consistent modelling, spatially resolved \CI{} observations, and deeper searches for \HCOp.}
  



   \maketitle
   \nolinenumbers
%
%-------------------------------------------------------------------

\section{Introduction}
Protoplanetary disks (PPDs) supply the material for planet formation and set the initial chemical and dynamical conditions of planetary systems. In the classical view, most protoplanetary disks disperse on timescales of only a few Myr because high-energy irradiation (EUV/FUV/X-rays) drives photoevaporative mass loss that clears the gas disks \citep{2001_Clarke,2001_Haisch,2010_Fedele,2014_Alexander,2015_Ribas}. Within this standard framework, the availability of primordial gas at ages of tens of Myr is generally not anticipated, and the transition from gas-rich PPDs to gas-poor debris disks is expected to occur within a few Myr.

Contrary to this standard picture, recent observations have revealed debris disks that contain not only dust but also detectable gas. A small but growing set of gas-rich debris disks around young A-type stars shows bright \CO{} emission \citep[e.g.,][]{Moor_2011,2013_Kospal,2017_Moor,DiFolco_2020,Higuchi_2020,Hales_2022,Rebollido_2022,Moor_2025,Szewczyk_2026}.

These systems probe the poorly understood interface between the end of the planet-forming epoch and the onset of collision-dominated debris evolution \citep{2018_Hughes}. In the conventional picture, a debris disk is sustained by collisions of planetesimals and contains little primordial gas; thus any detected gas is often interpreted as secondary, produced by ongoing release from solids \citep{2016_Marino, 2017_Matra, 2020_Marino}.
Because \CO{} is readily photodissociated by UV photons, steady secondary \CO{} requires continuous replenishment and/or effective shielding \citep{Kral_2016,2019_Hales}.
Neutral atomic carbon is particularly informative because it is a direct photoproduct of CO and is expected to have a high abundance in the secondary origin scenario \citep{2017_Higuchi,Kral_2017,Cataldi_2018,2019_Higuchib,2019_Higuchia,Kral_2019,Cataldi_2023,Brennan_2024}.

ALMA searches for commonly observed molecular tracers (e.g., \HCOp, \CN) in CO-rich debris disks around A stars have largely resulted in non-detections, even when comparable observations of evolved Herbig Ae protoplanetary disks do detect such species \citep{Matra_2018,Klusmeyer_2021,Smirnov-Pinchukov_2022}. These molecular upper limits can be used to study the origin of gas-rich debris disks.

In parallel, recent theoretical work has raised the plausibility that at least a subset of gas-rich debris disks can be long-lived remnants of protoplanetary disks that have become severely depleted in small dust grains \citep{Nakatani_2023,Ooyama_2025}. Theoretical disk evolution models showed that the observed demographics (predominantly A-type stars, $\sim$ a few tens of Myr) can be understood if small-grain depletion reduces FUV photoevaporation, allowing gas to survive much longer. 
\citet{Ooyama_2025} extended this picture using one-dimensional (1D) disk evolution simulations that incorporate stellar evolution and time-dependent photoevaporation, finding that gas survival beyond 10 Myr can occur across stellar masses for sufficiently massive initial disks, with the longest lifetimes robustly occurring around $2\Msun$ across a wide parameter space. This result is especially relevant because several of the most CO-rich debris disks are hosted by early A stars \citep{2017_Moor,2018_Hughes,Moor_2025}.

Together, these studies showed that an H$_2$-dominated gas reservoir can survive beyond 10 Myr, satisfying a necessary but not sufficient condition for the primordial-origin scenario. However, whether such long-lived disks can reproduce the observed \CO{} masses and other key chemical constraints has not been tested thermochemically. Here, we address this gap by assessing the chemical viability of the primordial-origin scenario using disk structures motivated by the long-lived models of \citet{Ooyama_2025} for a $2\,\Msun$ star. In particular, we focus on three issues: the survival of substantial \CO{} together with the low \CI/\CO{} ratios inferred from observations; the compatibility of the primordial-origin scenario with the observed non-detections of molecular ions such as \HCOp{}; and the physical conditions under which these constraints can be satisfied simultaneously, especially the roles of dust depletion, grain population, irradiation geometry, and cosmic-ray ionisation level.

To address these issues, we employ the thermochemical code \textsc{Cloudy} \citep{Ferland_2017,Chatzikos_2023} to compute the radiative transfer and chemical equilibrium of disk models representative of 20--40 Myr systems, thereby predicting the radial distributions and abundances of \CO, \CI, and \HCOp{} across the explored parameter space. By mapping how these observables respond to changes in dust content, irradiation conditions, and cosmic-ray ionisation rate, we identify which regions of parameter space remain compatible with current observational data, which remain problematic, and which future spatially resolved measurements would most clearly distinguish between primordial and secondary origins.

This paper is organised as follows. Sect.~\ref{sec:methods} describes the adopted disk structures, dust content, and irradiation setup. Sect.~\ref{sec:results} presents how \CO, \CI, and key molecular ions respond across the explored parameter space and also presents synthetic observables. In Sect.~\ref{sec:discussions}, we confront these models with resolved observations and with molecular non-detections in CO-rich debris disks, identifying which combinations of parameters remain compatible with the data. Sect.~\ref{sec:limitation} presents the main caveats of our approach. Sect.~\ref{sec:conclusions} summarises the resulting constraints on the primordial-origin scenario and discusses implications for the evolutionary pathways from protoplanetary disks to gas-rich debris disks.

\section{Methods}
\label{sec:methods}
We use gas surface-density and temperature profiles taken from the 1D disk-evolution simulations of \citet[][hereafter O25]{Ooyama_2025} as inputs to our thermochemical post-processing. We do not perform new disk-evolution simulations in this work. We focus on the model for a $2\,\Msun$ star (O25 ``FID-2''), motivated by the observational finding that CO-rich gas-bearing debris disks are preferentially hosted by early A-type stars. This model also represents the parameter region in which the primordial-origin scenario remains viable up to ages of tens of Myr. For completeness, we first summarise the O25 methodology used to generate the adopted profiles and then describe the thermochemical calculations performed in this section.

\subsection{1D disk evolution model}
The adopted profiles were generated with the long-term 1D disk-evolution model presented by O25. The O25 model includes EUV- and X-ray-driven photoevaporation by the central star and neglects FUV photoevaporation to represent a small-grain-depleted disk in which grain photoelectric heating is suppressed. The model also neglects the externally driven photoevaporation because a typical external radiation field ($\sim 1G_0$) does not lead to a high photoevaporation rate \citep{Winter_2018}. We assume a molecular hydrogen-dominated gas, similar to the interstellar medium (ISM).

The evolution of the disk's surface density is governed by
\begin{equation}
\begin{split}
\frac{\partial \Sigma_{\rm g}}{\partial t}
&=
\frac{1}{R}\frac{\partial}{\partial R}
\left[
\frac{2}{R\Omega_{\rm K}}
\left(
\frac{\partial}{\partial R}\left(R^{2}\Sigma_{\rm g}\alpha_{R\phi}c_{s}^{2}\right)
+
R^{2}(\rho c_{s}^{2})_{\rm mid}\alpha_{\phi z}
\right)
\right]\\
&\quad 
-C_{\rm w}(\rho c_{s})_{\rm mid}
-
\dot{\Sigma}_{\rm pw},
\label{eq:sigma_evol}
\end{split}
\end{equation}
where $\Omega_{\rm K}$ is the Keplerian angular velocity, $c_s$ the midplane sound speed, and $(\rho c_s)_{\rm mid}$ and $(\rho c_s^2)_{\rm mid}$ are evaluated at the disk midplane (see O25 for definitions and the derivation from the general form of \citealt{2016_Suzuki}). The first term represents radial mass transport driven by turbulent viscosity (parameterised by $\alpha_{R\phi}$) and the wind torque (parameterised by $\alpha_{\phi z}$), while the second and third terms describe mass loss by the MHD disk wind and by photoevaporation, respectively. We adopt the MRI-inactive prescription for the turbulent stress with $\alpha_{R\phi}=8\times10^{-5}$, which is consistent with the gas masses inferred for observed gas-rich debris disks \citep{Ooyama_2025}.

We choose the same initial surface density profile as O25 (the total initial disk mass satisfies $M_{\rm disk,0}/M_\star=0.1$).
The midplane temperature $T(R,t)$ is computed by combining irradiation and viscous heating
($T=\left[T_{\rm irr}^{4}+T_{\rm vis}^{4}\right]^{1/4}$; O25 Eqs.~19--21), including the stellar-evolution-dependent irradiation field adopted in O25. As a reference value, the adopted radial temperature profile gives $T=56\,\mathrm{K}$ at $R=100\,\mathrm{au}$.

\begin{figure}[h]
    \centering
    \includegraphics[width=1.\linewidth]{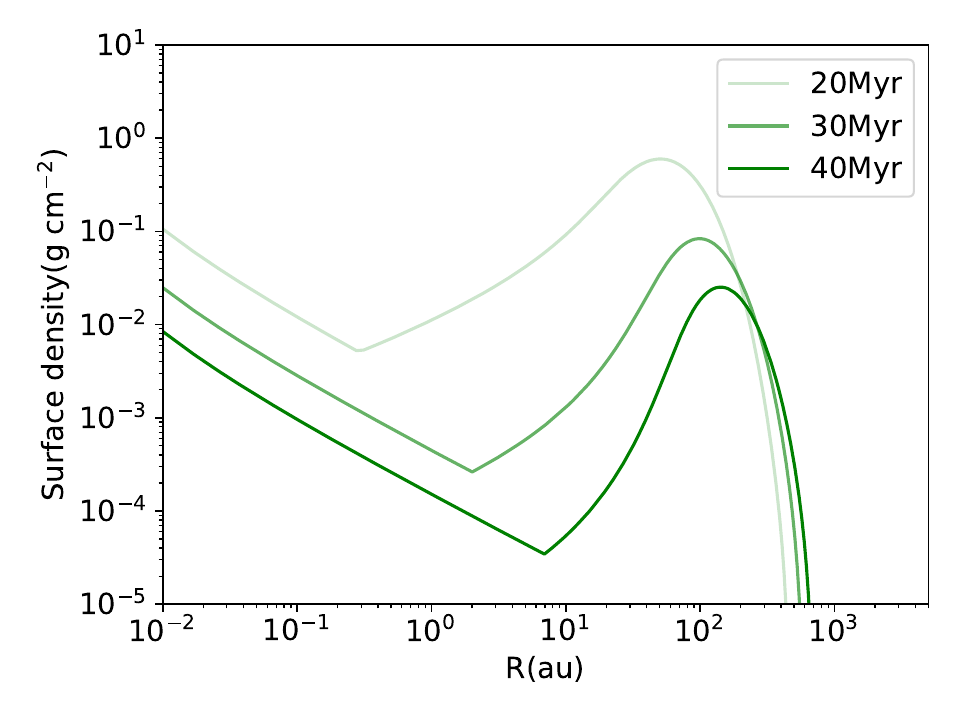}
    \caption{The evolution of the gas surface density of the adopted disk model around a 2$\,\Msun$ star. The profiles are taken from \citet{Ooyama_2025}. }
    \label{fig:hydro}
\end{figure}

Figure~\ref{fig:hydro} shows the corresponding temporal evolution of $\Sigma_{\rm g}(R,t)$ for the adopted model (data taken from O25).
These radial profiles of $\Sigma_{\rm g}(R)$ and $T(R)$ are then used as inputs to construct the axially symmetric two-dimensional (2D) density structure and to perform the thermochemical post-processing described in Sect.~\ref{sec:cloudy}.

\subsection{Post-processing via \textsc{Cloudy}}
\label{sec:cloudy}
We post-process the hydrodynamic disk profiles that provide the gas surface density $\Sigma_{\rm g}(R)$ and a representative temperature profile $T(R)$ as a function of cylindrical radius $R$ with \textsc{Cloudy} \citep{Ferland_2017,Chatzikos_2023} to calculate the radiative transfer and chemical equilibrium. We investigate both externally irradiated and internally irradiated disks.
In this section, we summarise the post-processing method via \textsc{Cloudy}.

A realistic disk is irradiated simultaneously by the host star and the interstellar radiation field (ISRF). Modelling their combined effects would require a multidimensional treatment because the relative importance of the two radiation sources depends on position within the disk. Instead, we consider two limiting cases separately: external irradiation by the ISRF and internal irradiation by the host star. We adopt the external-irradiation case as our fiducial model because the ISRF is expected to dominate CO photodissociation even around intermediate-mass stars \citep{Matra_2015}. The internal-irradiation models are used to assess the sensitivity of the chemistry to direct stellar irradiation. The limitations of this separation are discussed in Sect.~\ref{sec:irradiation_limitation}.

For the external-irradiation models, we calculate independent vertical columns at each cylindrical radius $R$. This external-irradiation setup does not include radially incident ISRF photons entering through the outer disk edge.

\begin{figure}[h]
    \centering
    \includegraphics[width=1.\linewidth]{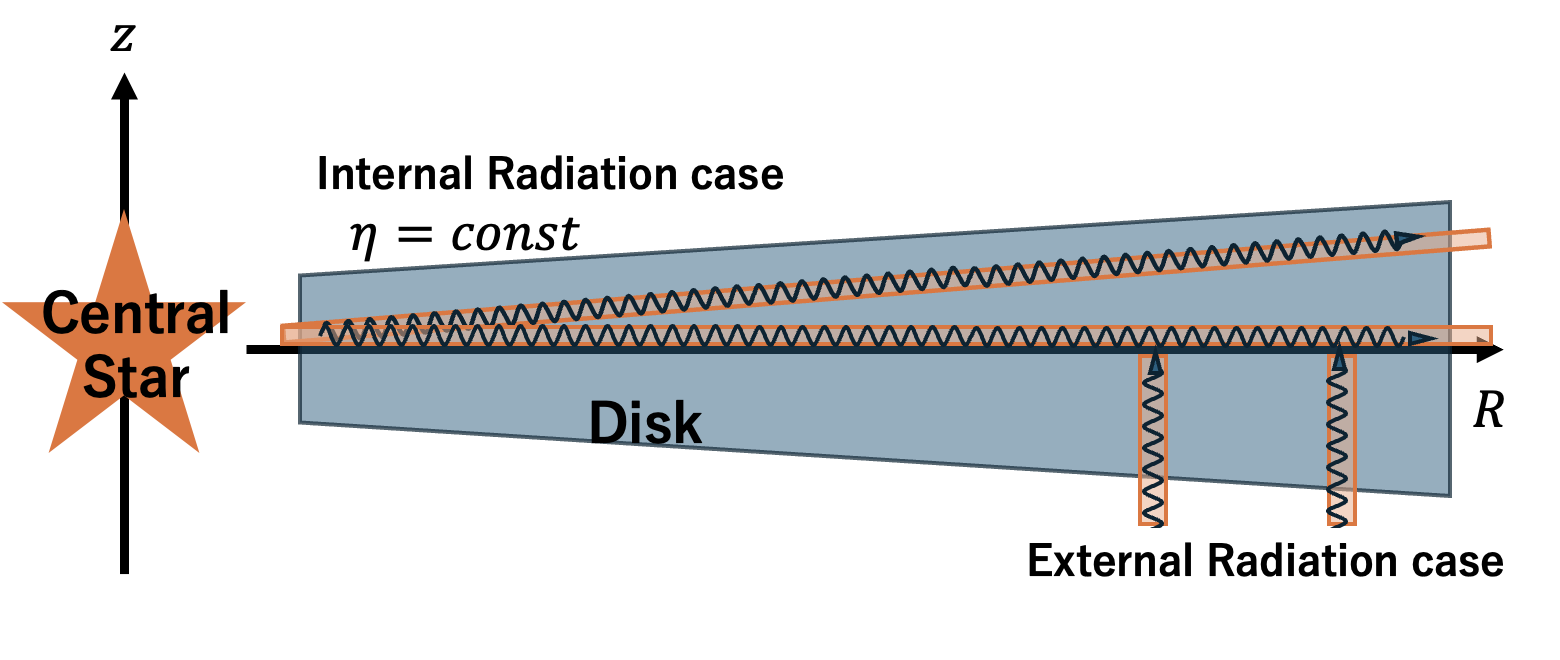}
    \caption{Schematic overview of the irradiation geometries explored in this work. In the internal-irradiation models, stellar radiation propagates radially outward from the central star. In the external-irradiation models, the ISRF is incident vertically on the disk surface.  }
    \label{fig:setting}
\end{figure}

For the internal-irradiation models, we calculate one-dimensional rays emerging radially from the central star. Figure~\ref{fig:setting} illustrates these two limiting geometries.

\subsubsection{Disk density and temperature structure from the disk evolution simulations}
\label{sec:disk_structure}

At each radius, we model the vertical density structure by a Gaussian profile,
\begin{equation}
\rho(R,z)=\rho_{0}(R)\exp\!\left[-\frac{z^{2}}{2H(R)^{2}}\right],
\label{eq:density}
\end{equation}
where $z$ is the height above the midplane and $H(R)$ is the pressure scale height computed assuming
vertical hydrostatic equilibrium and vertically isothermal structure,
\begin{equation}
H(R) = \frac{c_{\rm s}(R)}{\Omega_{\rm K}(R)}, \qquad
c_{\rm s}(R)=\sqrt{\frac{k_{\rm B}T(R)}{\mu m_{\rm H}}}, \qquad
\Omega_{\rm K}(R)=\sqrt{\frac{GM_\star}{R^{3}}}.
\label{eq:scale_height}
\end{equation}
We adopt a mean molecular weight $\mu=2.34$ and stellar mass $M_\star=2\,M_\odot$.
The midplane density $\rho_0$ is fixed by the requirement that the vertical column reproduces the
input surface density,
\begin{equation}
\rho_{0}(R)=\frac{\Sigma_{\rm g}(R)}{\sqrt{2\pi}\,H(R)}.
\label{eq:base_density}
\end{equation}
\textsc{Cloudy} uses the total hydrogen number density $n_{\rm H}$; we therefore convert the mass density to
\begin{equation}
n_{\rm H}(R,z)=\frac{\rho(R,z)}{1.4\,m_{\rm H}},
\label{eq:number_density}
\end{equation}
where the factor $1.4$ approximately accounts for helium and metals.

For the external-irradiation calculations, \textsc{Cloudy} follows
the upper half of the disk from the midplane to $z=7H(R)$. Assuming
reflection symmetry about the midplane, we calculate the full
two-sided vertical column density as
\begin{equation}
N_i(R)=2\int_0^{7H(R)} n_i(R,z)\,{\rm d}z.
\label{eq:column_density}
\end{equation}
The column densities shown in
Figs.~\ref{fig:fid_column} and \ref{fig:low_cr} include this factor of two and therefore represent the full vertical columns through both sides of the disk.

The disk-integrated mass is then computed as
\begin{equation}
M_{i}=2\pi\int_{R_{\rm in}}^{R_{\rm out}} \Sigma_{i}(R)\,R\,{\rm d}R,
\label{eq:Mi}
\end{equation}
where $R_{\rm in}$ and $R_{\rm out}$ are the inner and outer sampled radii of the adopted disk structure. We evaluate Eq.~\ref{eq:Mi} numerically by interpolating $\Sigma_i(R)$ over the sampled radii and integrating over $R$. For the internal radiation case, we first calculate the number density of species $i$, $n_i(r,\eta)$, along rays of constant height-to-radius ratio $\eta$, then reconstruct $n_i(R,z)$ using $r=\sqrt{R^2+z^2}$ and integrate over $z$ to obtain the surface density.
Using an independent midpoint-annulus integration changes most integrated masses by a few percent to about 15\%, which is smaller than the order-of-magnitude trends discussed below.

Along each vertical column or radial ray, the density is constructed from the adopted O25 surface-density profile using Eqs.~\ref{eq:density}--\ref{eq:number_density}. In our production runs, we do not solve for the gas temperature self-consistently in \textsc{Cloudy}. Instead, we adopt the temperature profile from the disk-evolution calculation of O25 and use \textsc{Cloudy} to compute the radiative transfer and chemical equilibrium on this prescribed thermal structure. The limitations associated with this temperature prescription are discussed in Sect.~\ref{sec:thermal_limitation}.

We also perform additional runs with \textsc{Cloudy}'s self-consistent thermal balance only as a sanity check. In the internal irradiation case, where stellar radiation is included, \textsc{Cloudy} predicts a midplane temperature of $T_{\rm mid}\simeq 40\,{\rm K}$ at $R\sim 100$~au. In contrast, in the external irradiation case, the predicted midplane temperature drops to $T_{\rm mid}\simeq 10\,{\rm K}$ at the same radius. At such low temperatures, CO freezes out onto dust grains and thermal desorption becomes negligible, shifting the dominant CO reservoir from the gas phase to CO ice. Therefore, for the external irradiation setup, adopting the temperature prescribed by the disk-evolution model (which effectively accounts for stellar heating) is likely more realistic for our system than using the \textsc{Cloudy} self-consistent temperature. We also verified that, in the internal irradiation case, recomputing the chemistry with \textsc{Cloudy} while fixing the temperature to the disk-evolution value changes the midplane CO gas density only mildly, typically by a factor of $\sim 2$.

For the isotopologue calculations, we assume a spatially constant elemental carbon isotope ratio of $^{12}{\rm C}/^{13}{\rm C}=30$. We do not include isotope-selective photodissociation or chemical isotope fractionation; instead, the $^{13}$CO abundance is obtained by scaling the CO isotopologue abundance according to the adopted isotope ratio. The main carbon isotope-exchange reaction becomes most effective at temperatures below approximately $\sim 30$\,K, whereas the reverse reaction suppresses strong fractionation at higher temperatures \citep{Woods_2009}. The adopted disk temperature is $56\,{\rm K}$ at $R=100$ au. We therefore expect chemical isotope fractionation to have a limited effect on the CO isotopologue ratio. This approximation may break down in the colder outer disk. The predicted $^{13}$CO masses and line ratios should therefore be interpreted as conditional on the adopted isotope ratio. Consequently, the predicted $^{13}$CO mass and the $^{12}$CO/$^{13}$CO line ratios depend directly on this assumed value and should be interpreted as model predictions conditional on $^{12}{\rm C}/^{13}{\rm C}=30$.

Note that in the disk midplane, the dominant formation and destruction rates of species are typically in the range $10^{-4}$--$10^{-11}{\rm ~s^{-1}}$ as shown in Fig.~\ref{fig:fid_reaction}. Even the slowest of these rates corresponds to a characteristic timescale of only $\sim 3\times10^3$ yr, which is several orders of magnitude shorter than the 20--40 Myr evolutionary timescale of the disk models. This indicates that the local abundances can adjust rapidly to the prescribed density, temperature, irradiation field, and ionisation rate, supporting the use of a quasi-steady chemical solution for these species in the midplane.

\subsubsection{External irradiation case}
\label{sec:external_field}

To model irradiation by an ISRF on the disk surface, we compute independent 1D vertical columns at radii $R=\{1,10,50,100,200,300,400,500,600\}\,\mathrm{au}$. However, for the 20 Myr case, the adopted surface density falls to zero at 500 and 600 au, and for the 30 Myr case, it becomes zero at 600 au as shown in Fig.~\ref{fig:hydro}. Therefore, we do not perform the calculations over those ranges.
For each $R$ we calculate the chemical structure from \textsc{Cloudy} spanning $z\in[0,7H(R)]$. We have checked that expanding the computational domain to $9H$ results in a change in CO density of only a few percent.

We include the ISRF using the built-in \texttt{table ISM} continuum in \textsc{Cloudy}, which is based on the local unattenuated ISRF of \citet{1987_Black}.
The \textsc{Cloudy} calculations were performed with version C23 \citep[][see also \citet{Ferland_2017}]{Chatzikos_2023}.

To facilitate a qualitative comparison with observations, we also compute synthetic line emission from the thermochemical models. For each 1D vertical slab at a given radius, we use the line fluxes and optical depths calculated by \textsc{Cloudy} for the relevant transitions. The radial line-flux profiles derived directly from the \textsc{Cloudy} outputs are shown without beam convolution. Beam convolution is applied only in the separate \textsc{radmc-3d} imaging calculation described in Sect.~\ref{sec:radmc3d}.

\subsubsection{Internal irradiation case}
\label{sec:internal_field}

To model irradiation by the central star, we also compute 1D rays emerging from the star and intersecting the disk
at constant height-to-radius ratio,
\begin{equation}
\eta \equiv \frac{z}{R},
\label{eq:zeta}
\end{equation}
with the ratio $\eta=\{0.00,0.05,0.10,0.20,0.30\}$. Each ray is discretised on a spherical radial coordinate $r$ (distance
from the star), which is mapped to cylindrical coordinates as
\begin{equation}
R(r)=\frac{r}{\sqrt{1+\eta^2}}, \qquad
z(r)=\frac{\eta\,r}{\sqrt{1+\eta^2}}.
\label{eq:r_zeta}
\end{equation}
At each location along the ray we evaluate $n_{\rm H}(R,z)$ using the Gaussian vertical structure (Eqs.~\ref{eq:density}--\ref{eq:number_density})
together with the input $\Sigma_{\rm g}(R)$ and $T(R)$. The resulting density and temperature profile is passed to \textsc{Cloudy}.

In the case of the internal radiation field, we include the host star spectrum as an input spectrum from ATLAS \citep{Kurucz_1991}. The stellar effective temperature $T_{\rm eff}$ and bolometric luminosity $L_\star$ are determined from the evolutionary calculations tabulated by \citet{2021_Kunitomo}. 
Early-type main-sequence A stars are expected to exhibit weak coronal X-ray/EUV emission because of the lack of an outer convective envelope, while cooler intermediate-mass stars (late-A/early-F) may show enhanced high-energy emission \citep{Schroder_2007,2018_Fossati}. We therefore neglect the intrinsic coronal XUV irradiation from the host A star in our fiducial model and focus on the photospheric UV field.

The internally irradiated models are not intended to replace the externally irradiated fiducial case. Rather, they provide the opposite limiting case in which stellar photons control the photochemistry along radial rays. These calculations are therefore useful for identifying which species are most sensitive to direct stellar irradiation, in particular \CI{} and \HCOp{} in the upper layers.

\subsubsection{Microphysics and numerical settings}
\label{sec:microphysics}

We include dust using either \textsc{Cloudy}’s built-in ISM grains or a user-supplied opacity table without small dust grains, scaled to the adopted DTG. In the case without small dust grains, we modified only the lower cutoff of the grain-radius distribution, adopting $a_{\min}=1\,\mu\mathrm{m}$, while retaining the default Cloudy distribution shape, dust abundance, optical constants, and size-resolved treatment for the remaining grain population.  We also disable stochastic grain heating (\texttt{no qheat}) for simplicity. We include a background cosmic-ray field and explore three cosmic-ray ionisation rates, $\zeta_{\rm CR}=10^{-17},\,10^{-18},\,10^{-19}\,\mathrm{s}^{-1}$. The cosmic microwave background is also included. We verified in test runs that enabling stochastic grain heating or turbulent line broadening does not change the chemical profiles.

\subsubsection{Synthetic $^{13}$CO observables with \textsc{radmc-3d}}
\label{sec:radmc3d}
To visualise how one representative primordial-remnant model would appear in current high-resolution observations, we post-process the thermochemical disk structure produced by \textsc{Cloudy} with a custom Python pipeline to generate the \textsc{radmc-3d} \citep{Dullemond_2012} line-transfer input. We use HD~121617 only as an illustrative observational reference and do not attempt an object-specific fit.

The \textsc{Cloudy} vertical profiles were mapped onto a regular spherical grid with $(n_r,n_\theta)=(60,500)$ assuming axisymmetry and mirror symmetry about the midplane. The resulting $n_{\tCO}$ and the gas temperature $T$ distributions are interpolated onto the 3D grid. A Keplerian azimuthal velocity field was adopted for a $2\,\Msun$ central star. Line excitation and transfer were computed in non-LTE using the escape-probability/LVG formalism implemented in \textsc{radmc-3d} (\texttt{lines\_mode=3}), with the escape length scale tied to twice the local thermal scale height and a small floor to avoid vanishing values.

Synthetic image cubes were produced for the \tCO{} transition with a field of view of $400~{\rm au}$ sampled by $400\times400$ pixels. We then rendered the model using one representative observational configuration, namely the inclination, position angle, angular resolution, and distance reported for HD~121617 (\citealt{Brennan_2026,MacManamon_2026}, inclination $i=44.1^\circ$, position angle $58.7^\circ$). The model cubes were converted from specific intensity to interferometric units by convolving each channel with an elliptical Gaussian beam matching the ALMA survey to resolve exo-Kuiper belt substructures (ARKS) synthesised beam ($0.13^{\prime\prime}\times0.12^{\prime\prime}$, PA $-72.6^\circ$), scaling to mJy\,beam$^{-1}$ using the beam solid angle, and adopting a distance of 117.9\,pc. For comparison, we also added Gaussian noise with an rms of $0.8~{\rm mJy\,beam^{-1}}$ and computed intensity maps by integrating over velocity (reported in mJy\,km\,s$^{-1}$\,beam$^{-1}$).

\subsubsection{Parameter space}
We focus on a $2\,\Msun$ host star because CO-rich gas-bearing debris disks are observed predominantly around early A-type stars and because this mass is favoured by the evolutionary models \citep{Ooyama_2025}. 
For the main grid, we vary the age, cosmic-ray ionisation rate, dust-to-gas mass ratio, and the irradiation geometry over all combinations of $t=20,30,40$ Myr,
$\zeta_{\rm CR}=10^{-17},10^{-18},10^{-19}\,{\rm s^{-1}}$,
and ${\rm DTG}=10^{-2},10^{-4}$.
We summarise the main parameter grid of this paper in Table~\ref{tab:param_space}. 

In addition to this grid, we perform one extreme dust-depletion experiment in Sect.~\ref{sec:plausibility}, in which only the DTG of the fiducial model is changed to ${\rm DTG}=10^{-8}$. 
This additional run is not included in the main grid because it is used only as a diagnostic test of the molecular-to-atomic hydrogen transition under extreme dust depletion.

\begin{table}
\caption{Explored parameter space of the \textsc{Cloudy} post-processing models.}
\label{tab:param_space}
\centering
\begin{tabular}{lc}
\hline\hline
Parameter & Values \\
\hline
Irradiation geometry & external, internal \\
Age [Myr] & 20, 30, 40 \\
CR ionisation rate $\zeta_{\rm CR}$ [$\mathrm{\,s^{-1}}$] & $10^{-17},\,10^{-18},\,10^{-19}$ \\
Dust-to-gas mass ratio (DTG) & $10^{-2},\,10^{-4}$ \\
Dust size distribution & with / without small grains \\
Gas-phase abundances & solar \\
Grain model & ISM grains \\
\hline
\end{tabular}
\end{table}

\section{Results}
\label{sec:results}
In this section, we examine how the chemical structure of long-lived primordial-remnant disks depends on the adopted physical and chemical parameters. 
We begin with a fiducial model with external irradiation, an age of 40 Myr, a dust-to-gas ratio of \(10^{-4}\), and the standard cosmic-ray ionisation rate \(\zeta_{\rm CR}=10^{-17}\,{\rm s^{-1}}\). We adopt this model as our fiducial case because its total disk mass is closest to the observed range inferred for gas-rich debris disks, while the $t=20$ and $30\,\mathrm{Myr}$ models are used to trace the temporal evolution.
Then we present how the profiles change with the cosmic-ray ionisation rate in Sect.~\ref{sec:cr_ionisation}, dust properties (DTG, size distribution) in Sect.~\ref{sec:dust}, and radiation field geometry in Sect.~\ref{sec:internal}. 
We also summarise the time evolution of disk-integrated masses of different species (Sect.~\ref{sec:evolution}).

\subsection{Radial distributions of species}
\label{sec:fid}

We first examine the externally irradiated disk model at $t=40\,\mathrm{Myr}$, adopting the standard cosmic-ray ionisation rate, $\zeta_{\rm CR}=10^{-17}\,\mathrm{s}^{-1}$, and a low dust-to-gas ratio, $\mathrm{DTG}=10^{-4}$. This setup corresponds to the dust-poor $2\,\Msun$ disk presented in the theoretical study of \citet{Ooyama_2025}, here extended to include external irradiation. Our aim is to investigate the chemical structure of such a disk as a possible analogue of a gas-rich debris disk.

\begin{figure}
    \centering
    \includegraphics[width=1.\linewidth]{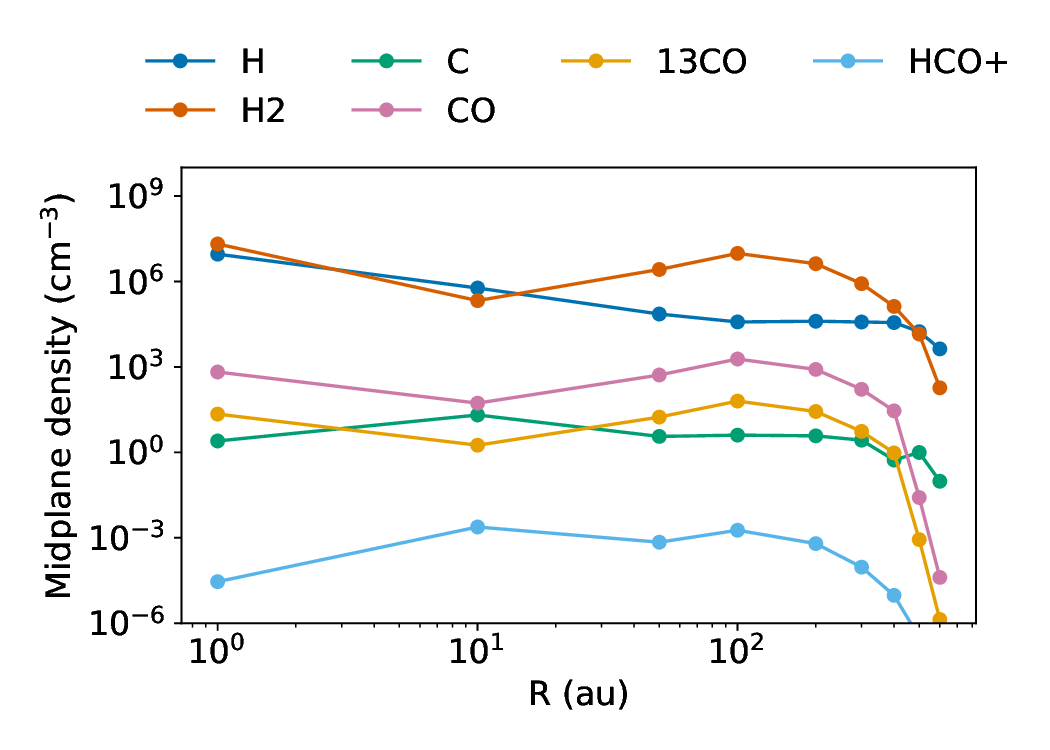}
    \caption{Midplane number-density profiles at $t=40\,\Myr$ for the fiducial externally irradiated model
(DTG$=10^{-4}$, $\zeta_{\rm CR}=10^{-17}\,\mathrm{s}^{-1}$).
Shown are H, H$_2$, CO, \tCO, \CI, and \HCOp{} number densities evaluated at $z=0$.}
    \label{fig:fid_external}
\end{figure}

Figure~\ref{fig:fid_external} shows the midplane number-density profiles of our fiducial disk. The underlying 1D disk-evolution calculation assumes a small-grain-depleted disk without FUV-driven photoevaporation and is not designed to reproduce the outer radius or total gas mass of any individual system. This fiducial model should therefore be regarded primarily as a chemically representative late-stage primordial-remnant disk, rather than as a unique fit to any individual system.

\begin{figure}
    \centering
    \includegraphics[width=1.\linewidth]{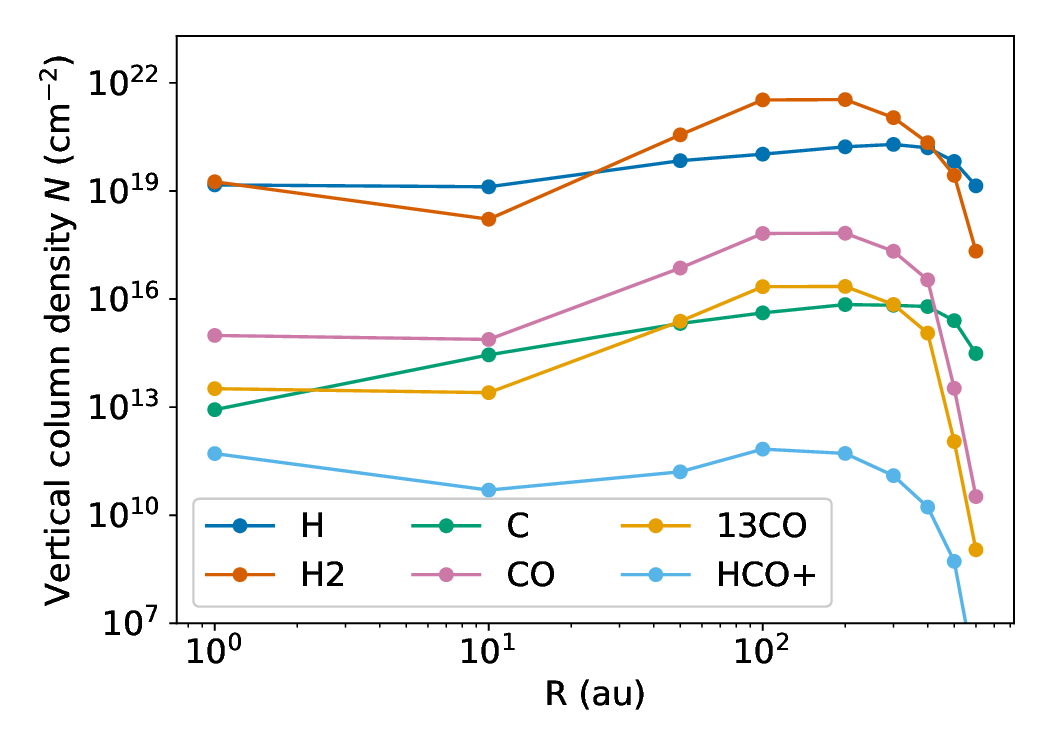}
    \caption{Full two-sided vertical column densities $N(R)$ of different species for the fiducial externally irradiated model.
Column densities are obtained by integrating the \textsc{Cloudy} vertical structure from the midplane to $z=7H(R)$.}
    \label{fig:fid_column}
\end{figure}

Throughout most of the disk midplane, hydrogen is predominantly molecular and provides the dominant gas reservoir.
CO is abundant around $\sim 100-400$ au, where $n(\mathrm{CO})$ is $\sim10^{2}$--$10^{3}\ \mathrm{cm^{-3}}$,
but it decreases sharply at larger radii as shielding weakens and chemical processing shifts in the tenuous outer disk.
The ratio of neutral atomic carbon to CO $n(\CI)/n(\CO)$ is low at small radii but rises toward the outer disk, reaching its maximum around the outermost region.
$n(\HCOp)$ shows a peak at intermediate radii ($R\sim100$ au) and becomes negligible in the outermost disk.
Figure~\ref{fig:fid_column} shows the column densities of the fiducial external irradiation case. The column densities of molecular species reach a maximum around $R\sim200\mathrm{\, au}$ and decrease in the outermost region due to the weak shielding, while the column densities of neutral atoms are higher in the outermost region.

Recent ARKS observations \citep{Brennan_2026} have suggested that, in some gas-rich debris disks, both $^{12}$CO and $^{13}$CO may be optically thick. We therefore also investigated the optical depths of the $^{12}$CO and $^{13}$CO lines in our models.
The $^{12}$\CO\ $J\!=\!3$--$2$ and $^{13}$\CO\ $J\!=\!3$--$2$ lines are optically thick ($\tau_{\rm midplane}>1$) in the vertical direction over $R\simeq10$--$300\,\mathrm{au}$.
Consequently, the \CO{} mass residing in regions with $\tau_{\rm midplane}<1$ accounts for less than $1\%$ of the total \CO{} mass and less than $10\%$ of the total $^{13}$\CO{} mass ($M_{{^{13}\CO},\mathrm{total}} = 9.9\times10^{-3} \Mearth,\, M_{{^{13}\CO},\mathrm{\tau<1}} \sim 0.5\times10^{-3} \Mearth$). We also confirm that the $^{12}$\CO\ $J\!=\!2$--$1$ line is similarly optically thick over $R\simeq10$--$300\,\mathrm{au}$.

To identify the dominant chemical pathways, Fig.~\ref{fig:fid_reaction} summarises the main formation and destruction
rates in the midplane for the fiducial model.
CO is sustained by a balance between gas-phase formation and ion-driven destruction,
while \ion{C}{i} is mainly produced by the dissociation of carbon-bearing molecules (including CO and minor C-molecules)
and is removed through reformation into CO and other molecules.
For HCO$^{+}$, the midplane abundance is controlled primarily by ion--molecule reactions initiated by cosmic-ray ionisation related to the availability of \Htp{}.
\begin{equation}
\label{eq:HCOp_formation}
    \mathrm{H_3^+ + CO \rightarrow HCO^+ + H_2}
\end{equation}
is among the dominant formation routes at the radii where \HCOp{} peaks. This is consistent with previous studies of chemistry in the interstellar medium and protoplanetary disks \citep{Herbst_1973,Leemker_2021}.
Since \(\mathrm{H_3^+}\) is initiated by cosmic-ray ionisation, the midplane \HCOp{} abundance is expected to be sensitive to the ionisation degree and hence to \(\zeta_{\rm CR}\). 
This motivates the parameter study of the ionisation rate presented below, where its effect is compared with those of dust abundance, grain size, irradiation geometry, and age.

\subsection{Dependence on the cosmic-ray ionisation rate}
\label{sec:cr_ionisation}
\begin{figure}
\centering
\includegraphics[width=1.\linewidth]{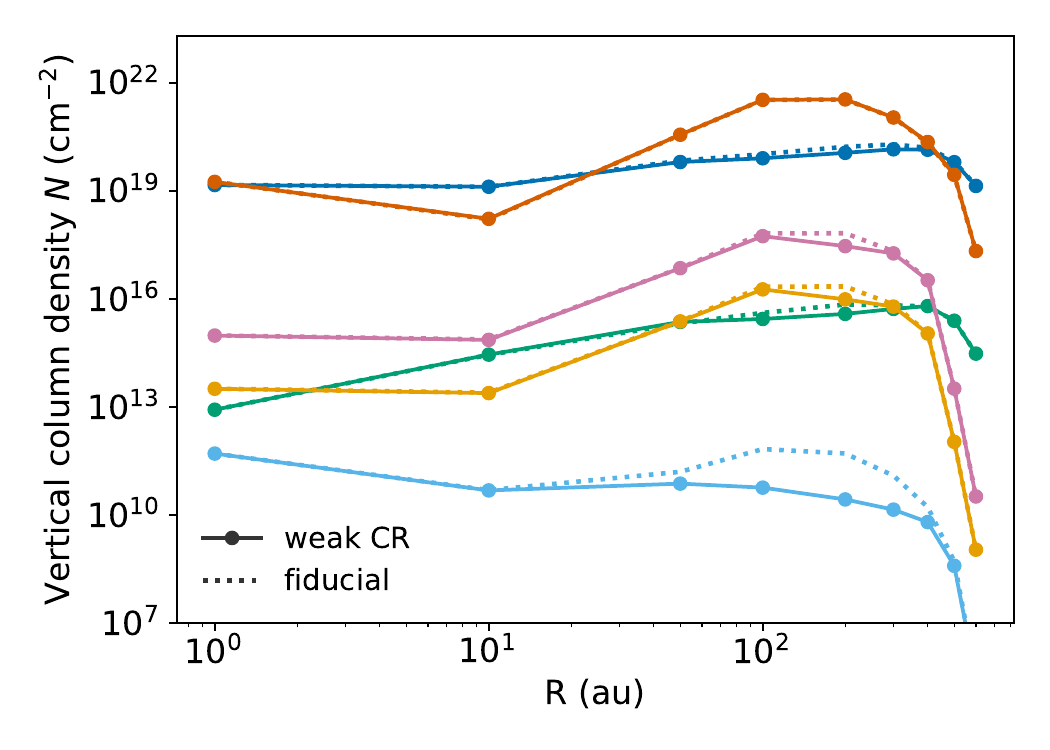}
\caption{Same as Fig.~\ref{fig:fid_column}, but comparing the weak-CR model with $\zeta_{\rm CR}=10^{-19}\,\mathrm{s}^{-1}$ and the fiducial model, shown by solid and dotted lines, respectively.
}
\label{fig:low_cr}
\end{figure}

We next examine how the adopted cosmic-ray ionisation rate affects the chemical structure.
The ionisation rate in gas-rich debris disks is poorly constrained, and studies of protoplanetary disks suggest that cosmic rays can be strongly attenuated in some environments \citep{Cleeves_2013,Cleeves_2015,Fujii_2022}.
We therefore compare the standard value, $\zeta_{\rm CR}=10^{-17}\,\mathrm{s}^{-1}$, with a suppressed value, $\zeta_{\rm CR}=10^{-19}\,\mathrm{s}^{-1}$.
Figure~\ref{fig:low_cr} shows the column-density profiles of the externally irradiated, low-DTG model with the suppressed cosmic-ray ionisation rate.
Compared with the fiducial model, lowering $\zeta_{\rm CR}$ has only a modest effect on the bulk molecular and atomic carbon reservoirs, while it strongly suppresses \HCOp{}.
The \CO{} and \CI{} column-density profiles remain broadly similar to those in the standard-CR case, whereas the \HCOp{} abundance decreases in the outer disk.
The resulting total \HCOp{} mass is reduced by  1--1.5 orders of magnitude, reaching $\sim2\times10^{-8}\,\Mearth$ in the weak-CR models.
The change in \HCOp{} abundance reflects a change in the dominant ion--molecule chemistry.
For the standard cosmic-ray ionisation rate, as in Eq.~\ref{eq:HCOp_formation}, \HCOp{} is mainly formed through \Htp{} ultimately produced by cosmic-ray ionisation.
When $\zeta_{\rm CR}$ is reduced, this pathway becomes inefficient and alternative channels, such as
\begin{equation}
\mathrm{H_2 + CO^+ \rightarrow HCO^+ + H},
\end{equation}
become relatively more important.
Thus, varying $\zeta_{\rm CR}$ selectively suppresses the molecular-ion chemistry without strongly changing the total CO and \CI{} reservoirs.

\begin{figure}[t]
    \centering
    \includegraphics[width=1.\linewidth]{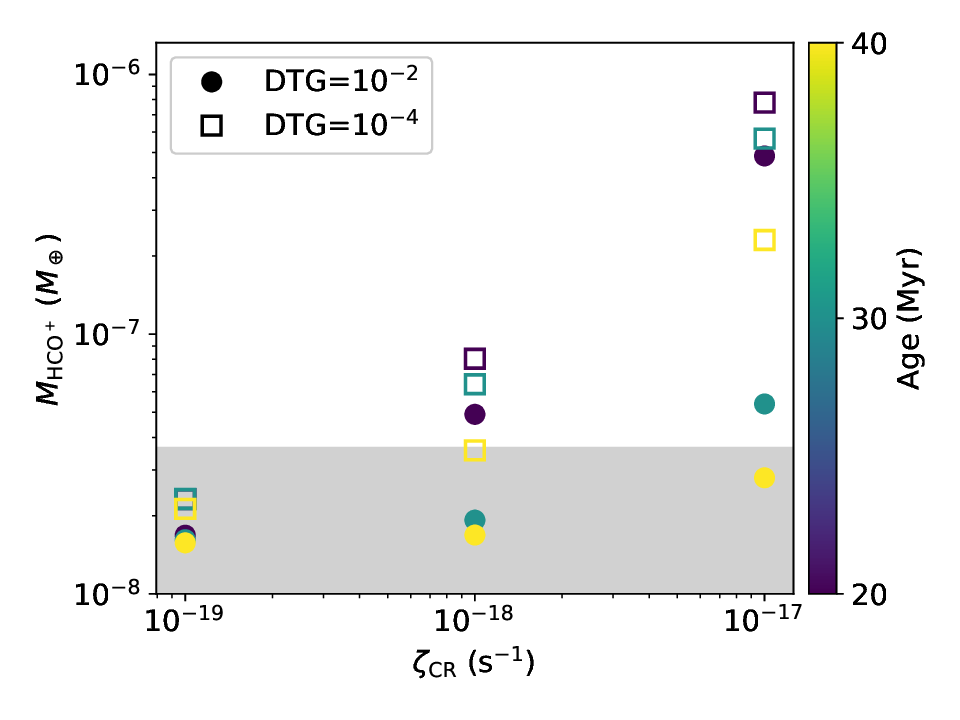}
    \caption{
        Total mass of \HCOp\ as a function of the cosmic-ray ionisation rate, $\zeta_{\rm CR}$, for the external irradiation models.
        Marker shape indicates the dust-to-gas ratio (DTG): filled circles correspond to ${\rm DTG}=10^{-2}$, while open squares correspond to ${\rm DTG}=10^{-4}$.
        The marker colour encodes the disk age (20, 30, and 40~Myr; see the colour bar). The gray shaded region marks the mass range below the largest LTE \HCOp{} upper limit inferred from the \HCOp{} $J=2\rightarrow1$ non-detections of \citet{Smirnov-Pinchukov_2022}; the source-specific limits span $6.7\times10^{-9}$--$3.7\times10^{-8}\,\Mearth$ for $T_{\rm ex}=20$--$50\,\mathrm{K}$.
        }
\label{fig:mhcop_zeta}
\end{figure}

Assuming optically thin LTE emission, the \HCOp{} mass of observed disks is
\begin{equation}
M_{\HCOp} =
\frac{4\pi d^2 F}{h\nu A_{ul}}
\frac{Q(T)}{g_u}
\exp(E_u/kT)
\,m_{\HCOp},
\end{equation}
where $d$ is the distance to the object, $F$ is the line energy flux, $\nu$ is the transition frequency, $A_{ul}$ is the Einstein coefficient, $Q(T)$ is the partition function, $g_u$ is the statistical weight of the upper level, $E_u$ is the upper-level energy, and $m_{\HCOp}$ is the mass of one \HCOp{} molecule.
Using the observational upper limits for gas-rich debris disks \citep{Smirnov-Pinchukov_2022}, this corresponds to a typical upper limit of a few $\times10^{-8}\,\Mearth$ for excitation temperatures of $T=20$--$50\,\mathrm{K}$. The \HCOp{} masses predicted by our weak-CR simulations are comparable to this limit, within a factor of two, despite the fact that no fitting was performed for individual sources.
Figure~\ref{fig:mhcop_zeta} summarises the resulting dependence of the total \HCOp{} mass on $\zeta_{\rm CR}$ for the externally irradiated models.

For the observational comparison, we also calculated the \HCOp{} $J=2\rightarrow1$ line luminosity predicted directly from the \textsc{Cloudy} emergent fluxes. The $3\sigma$ integrated-flux upper limits of 20.1--31.6~mJy\,km\,s$^{-1}$ for the four CO-rich debris disks observed by \citet{Smirnov-Pinchukov_2022} correspond to line-luminosity limits of $1.1\times10^{23}$--$3.6\times10^{23}\,\mathrm{erg\,s^{-1}}$.  Across the weak-CR disks, the predicted luminosities decrease to $1.6\times10^{23}$--$3.4\times10^{23}\,\mathrm{erg\,s^{-1}}$, overlapping the observed upper-limit range. The standard-CR, low-DTG models predict $L_{\HCOp}=3.4\times10^{24}$--$7.3\times10^{24}\,\mathrm{erg\,s^{-1}}$ and therefore clearly exceed these limits. The weak-CR models do not fall below every source-specific limit, because the present disk structures are not fitted to the individual systems. Within the explored external-irradiation framework, the non-detections therefore favour ionisation rates below the standard interstellar value in the primordial-origin scenario, with $\zeta_{\rm CR}\sim10^{-19}\,\mathrm{s}^{-1}$ providing the closest agreement.

\subsection{Effect of dust-to-gas mass ratio and dust size distributions}
\label{sec:dust}
We next investigate the DTG dependence of the chemical profiles of gas-rich debris disks. 
In the primordial-origin scenario, a low DTG is expected because the lack of FUV-driven photoevaporation is required for long-lived gas-rich disks and radiative pressure may remove dust particles \citep{2008_Wyatt,Krivov_2010}. However, both the timing and the degree of dust depletion remain uncertain. Therefore, we also perform the calculations for higher ISM-like DTG values ($10^{-2}$). The high-DTG models are not intended as self-consistent evolutionary models. They are chemistry-only sensitivity experiments performed on the same gas surface-density background to isolate the effect of dust abundance and grain surface area.

\begin{figure}
    \centering
    \includegraphics[width=1.\linewidth]{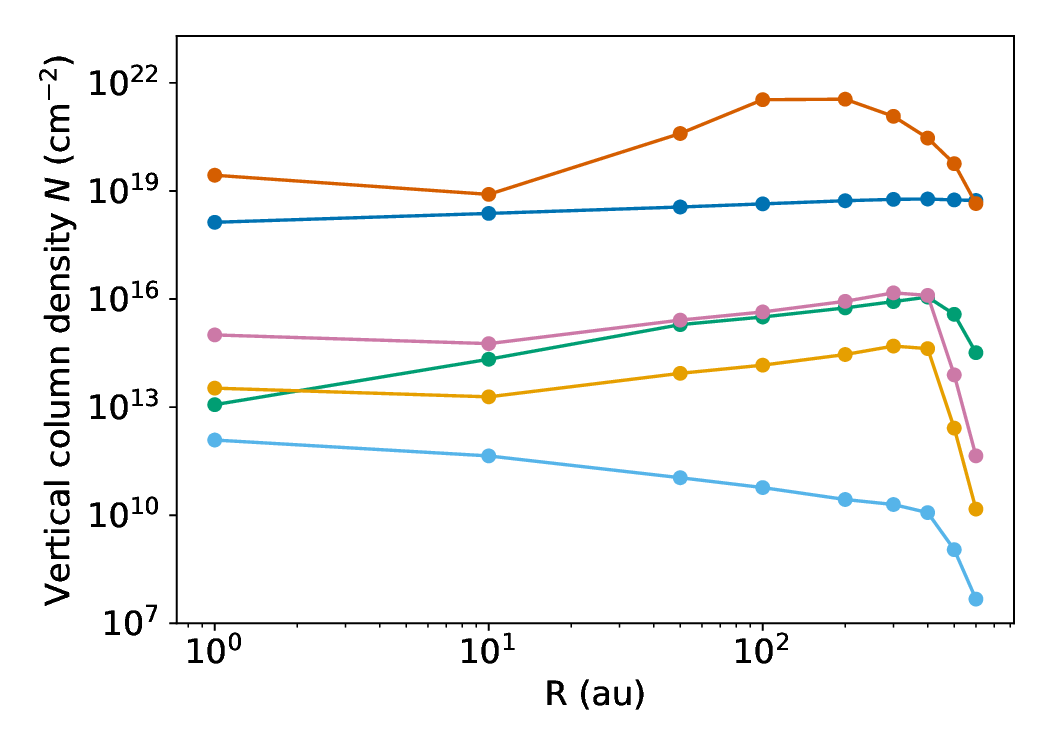}
    \caption{Same as Fig.~\ref{fig:fid_column} but for ISM-like dust (DTG$=10^{-2}$).  }
    \label{fig:high_DTG_external}
\end{figure}

Figure~\ref{fig:high_DTG_external} shows the radial profiles of column density for models with high DTG.
In the inner disk, grain-related processes dominate the \CO{} chemistry in both cases.
The \CO{} budget is primarily regulated by the exchange between the gas phase and grain surfaces.

For the dust-poor disks, the contribution of grain-related reactions is strongly reduced in the outer disk.
In particular, CO formation via 
\begin{equation}
\mathrm{CH_2} + \mathrm{O} \rightarrow \mathrm{CO} + 2\mathrm{H},
\end{equation}
becomes dominant, while 
\begin{equation}
\mathrm{He^+} + \mathrm{CO} \rightarrow \mathrm{C^+} + \mathrm{O} + \mathrm{He},
\end{equation}
dominates the destruction pathways in the case of ISM-like DTG.

In the dust-rich model, grain-related reactions remain efficient at large radii and remove CO from the gas phase, reducing the outer-midplane CO abundance. By contrast, in the dust-poor model, the reduced grain surface area weakens these removal channels and allows a larger fraction of carbon to remain in gas-phase molecules.

In gas-rich debris disks, small grains can be depleted by radiation pressure. We therefore vary the dust size distribution to assess the effects of small-grain depletion.

\begin{figure}
    \centering
    \includegraphics[width=1.\linewidth]{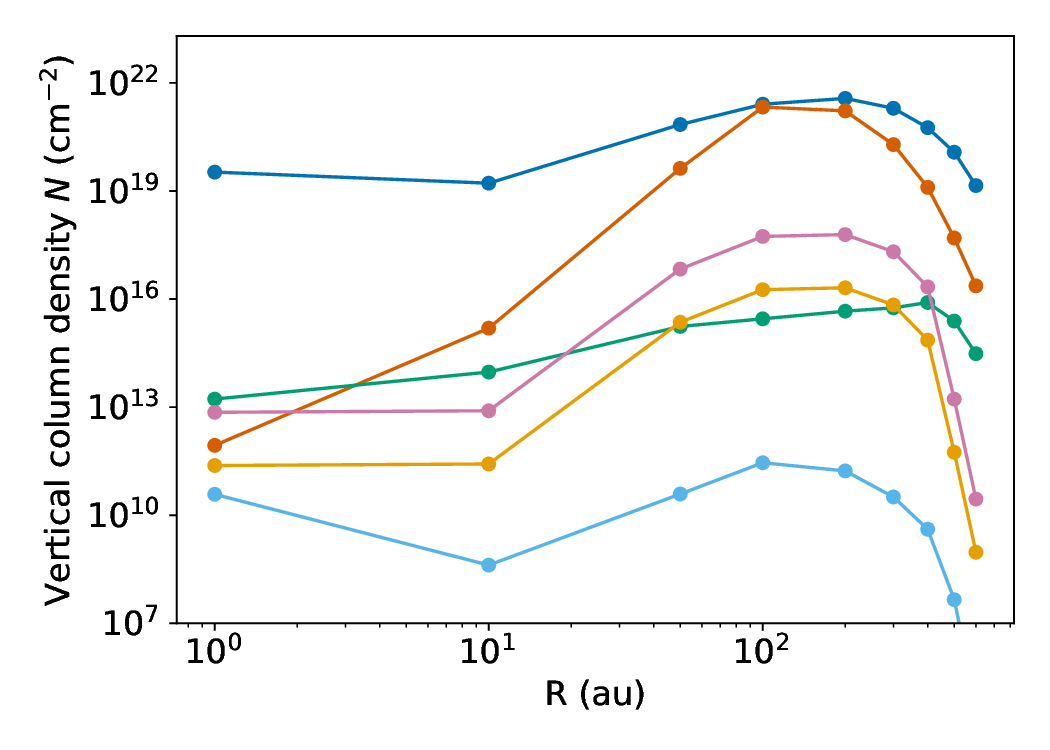}
    \caption{Same as Fig.~\ref{fig:fid_column} but adopting a grain population depleted of small grains (minimum grain size $a_{\min}=1\,\mu$m).}
    \label{fig:large_grain_external}
\end{figure}
Figure~\ref{fig:large_grain_external} shows the column density profiles in the case of large grains ($a_{\min}=1\,\mu$m). 
The CO column density decreases due to the elimination of small dust particles and the reduction in dust shielding.
However, the depletion of small grains does not remove the molecules entirely, and gas-phase CO survives.

The DTG and grain-size experiments point to the same general trend: reducing the effective surface area of small grains weakens grain-mediated processing and allows a larger fraction of carbon to remain in the gas phase. However, the effect is not monotonic for all species. While a lower small-grain content helps to preserve gas-phase CO in the outer disk by reducing grain-related removal channels, it also reduces dust shielding and can therefore lower the molecular column densities when the depletion of small grains becomes too strong. The resulting behaviour is thus a balance between reduced grain chemistry and reduced attenuation of the radiation field.

\subsection{Effect of stellar irradiation: internal-irradiation case}
\label{sec:internal}
\begin{figure*}[h]
    \centering
    \includegraphics[width=1.\linewidth]{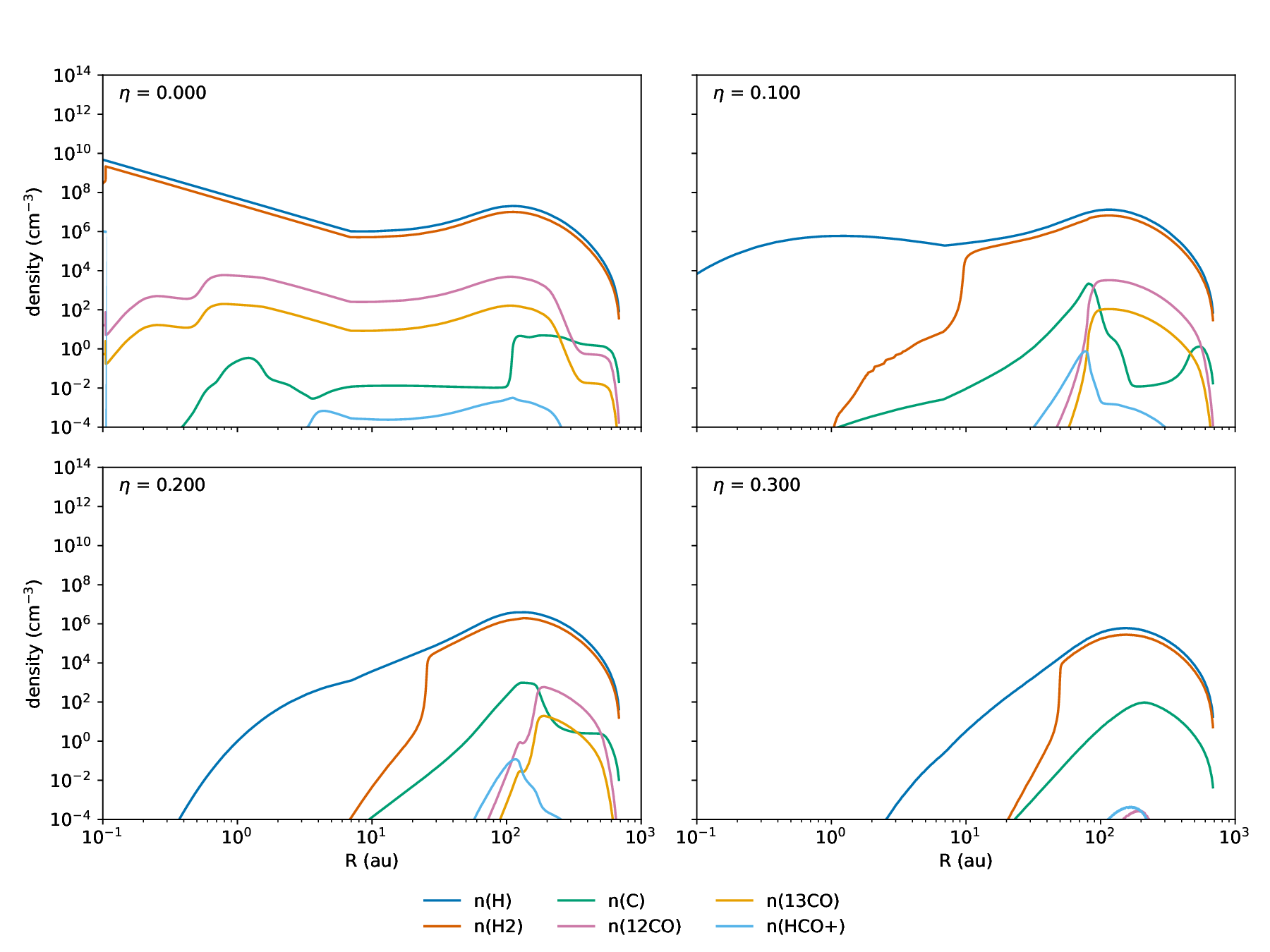}
    \caption{Radial profile for the internally irradiated model at $t=40\,\mathrm{Myr}$ (DTG$=10^{-4}$), computed along rays of constant height-to-radius ratio $\eta=z/R$. Different panels show $\eta=0.0, 0.1, 0.2, 0.3$.}
    \label{fig:fid_internal_density}
\end{figure*}

The internally irradiated models are used as a bracketing experiment to assess the sensitivity of the chemistry to direct stellar photons, especially in the inner disk and in the upper layers. 
In the case of internal irradiation, we calculate the density profile for each height-to-radius ratio $\eta$.
We show the radial profile in Fig.~\ref{fig:fid_internal_density}.
The densities tend to be larger in the midplane, but \HCOp{} and \CI{} tend to be larger in the slightly upper layers ($\eta=0.1$). \HCOp{} is formed via \CHtp{} and \CI{} is formed via photodissociation, and these chemical reaction rates are higher in this region.

Photodissociation significantly reduces the molecular abundance in the inner region; however, because most of the mass resides at larger radii, the disk-integrated masses are insensitive to this inner-disk photodissociation. In the internally irradiated models, the total \HCOp{} mass exceeds the observational upper limit by about two orders of magnitude. In the externally irradiated models, by contrast, \HCOp{} is suppressed in the upper layers and the total mass remains compatible with the observations.

Photodissociation strongly affects the internally irradiated models, but its quantitative effect is sensitive to the adopted CO photodissociation prescription. We discuss the photodissociation treatment in Sect.~\ref{sec:limitation}. 

\subsection{Evolution of total gas mass of different species}
\label{sec:evolution}
In this section, we summarise the age dependence of the disk-integrated gas masses between 20 and 40\,Myr. 
The radial distributions of the main species change only modestly in shape over this interval, while their normalisations evolve with the declining gas surface density. 
We therefore quote the integrated masses directly rather than introducing an additional figure for the nearly self-similar radial profiles (see Table~\ref{tab:species_masses}).

Over this interval, the gas budget is dominated by molecular hydrogen, while carbon-bearing species evolve more rapidly. The molecular hydrogen mass decreases monotonically from $\sim6\times10^{2}\,\Mearth$ at 20\,Myr to $\sim2\times10^{2}\,\Mearth$ at 30\,Myr and $\sim10^{2}\,\Mearth$ at 40\,Myr. 
Note that the absolute masses are set primarily by the adopted O25 surface-density profiles and are therefore conditional on the assumed small-grain-depleted evolutionary history; the thermochemical post-processing does not independently determine the total gas reservoir.

The CO mass drops from $\sim1.7\,\Mearth$ at 20\,Myr to $\sim0.5\,\Mearth$ at 30\,Myr and $\sim3\times10^{-1}\,\Mearth$ at 40\,Myr, corresponding to a factor of fives reduction across 20\,Myr. In contrast, atomic hydrogen mass increases with age by a factor of 2. As in Sect.~\ref{sec:fid}, since a large fraction of the CO mass is concentrated in the optically thick region, CO masses inferred under an optically thin assumption may underestimate the true CO masses of these disks.

Neutral atomic carbon stays at a low level of $\sim10^{-3}\,\Mearth$ because the atomic carbon is formed via photodissociation of CO. As a consequence, the integrated mass ratio $M_{\CI}/M_{\CO}$ increases with age. Finally, $M_{\HCOp}\sim10^{-6}\,\Mearth$ at 20\,Myr and decreases gradually to $\sim10^{-7}\,\Mearth$ at 40\,Myr.

Reducing the cosmic-ray ionisation rate mainly affects \HCOp{} and leaves the other major species nearly unchanged. In the weak-CR models, the total \HCOp{} mass decreases by factors of $\sim11$--34 in the low-DTG models, and all the weak-CR models yield $M_{\mathrm{HCO}^+}\sim2\times10^{-8}\,\Mearth$. Under the standard cosmic-ray ionisation rate, comparably low \HCOp{} masses are reached only in models with relatively high dust-to-gas ratios and at later evolutionary stages. This selective response explains why \HCOp{} provides the strongest observational constraint on the ionisation environment of primordial-remnant disks. Taken together, these results suggest that the combination of external-like irradiation, low small-grain content, and weak ionisation provides the most promising route within our explored primordial-remnant models for reproducing the observed combination of strong \CO, weak \CI, and non-detected \HCOp.

\subsection{Line fluxes and synthetic image}
To facilitate comparison with spatially resolved observations, we compute synthetic line observables for a representative weak-CR model that lies closest to the global chemical constraints identified above, namely the 40 Myr, low-DTG, weak-CR case. Our aim here is not to fit any individual object in detail, but to assess whether a primordial-remnant model with substantial CO, weak \CI{}, and low \HCOp{} luminosity also yields observational properties broadly similar to those inferred for resolved gas-rich debris disks. For the image synthesis, we adopt one representative observational setup of HD~121617 as a convenient fiducial case. This model is not intended as a best fit to HD~121617; the system is used only to define a representative viewing geometry, distance, and beam size for illustration.

\begin{figure}[h]
    \centering
    \includegraphics[width=1.\linewidth]{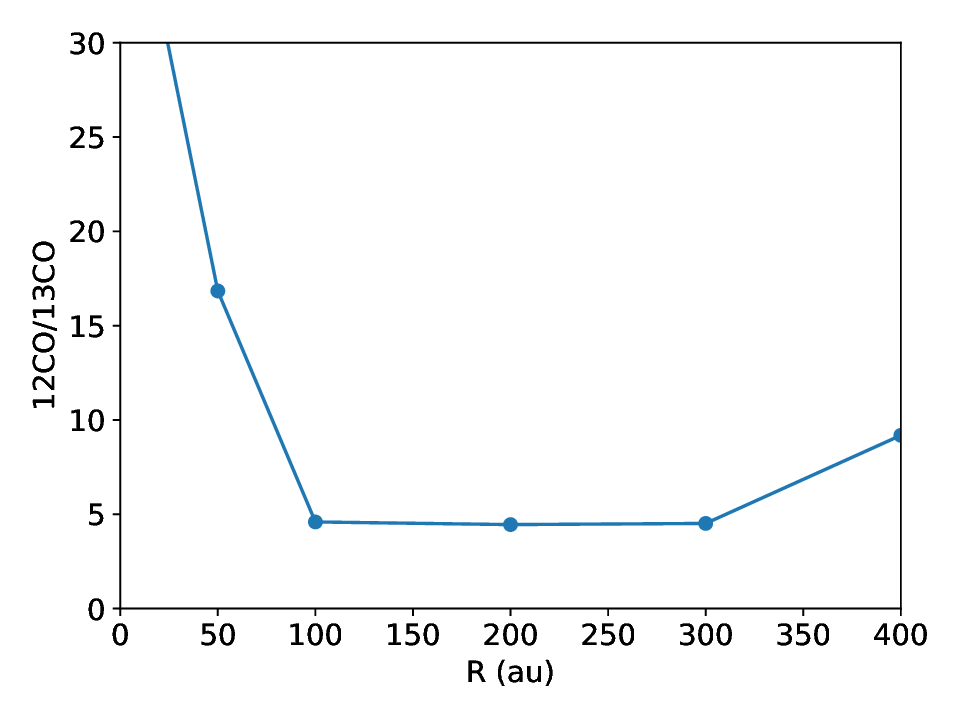}
    \caption{Radial profile of the CO line-flux ratio $F(^{12}\mathrm{CO}\;J\!=\!3\!-\!2)/F(^{13}\mathrm{CO}\;J\!=\!3\!-\!2)$ in the model whose total mass best matches the observations (40 Myr, weak CR, and low DTG).}
    \label{fig:low_CR_CO_ratio}
\end{figure}

Around $R\sim100$~au, the radially resolved isotopologue ratio in the $J=3$--2 transition remains low, with $^{12}$CO/$^{13}$CO $\simeq 5$ as shown in Fig.~\ref{fig:low_CR_CO_ratio}. This behaviour is consistent with the ARKS results for HD~121617, where the $^{12}$CO/$^{13}$CO integrated-intensity ratio is $\sim 2$ at the $^{12}$CO peak and stays approximately flat, with only a modest increase to $\sim 3$ between $\sim 60$ and $90$~au \citep{MacManamon_2026}. 

\begin{figure}[h]
    \centering
    \includegraphics[width=1.\linewidth]{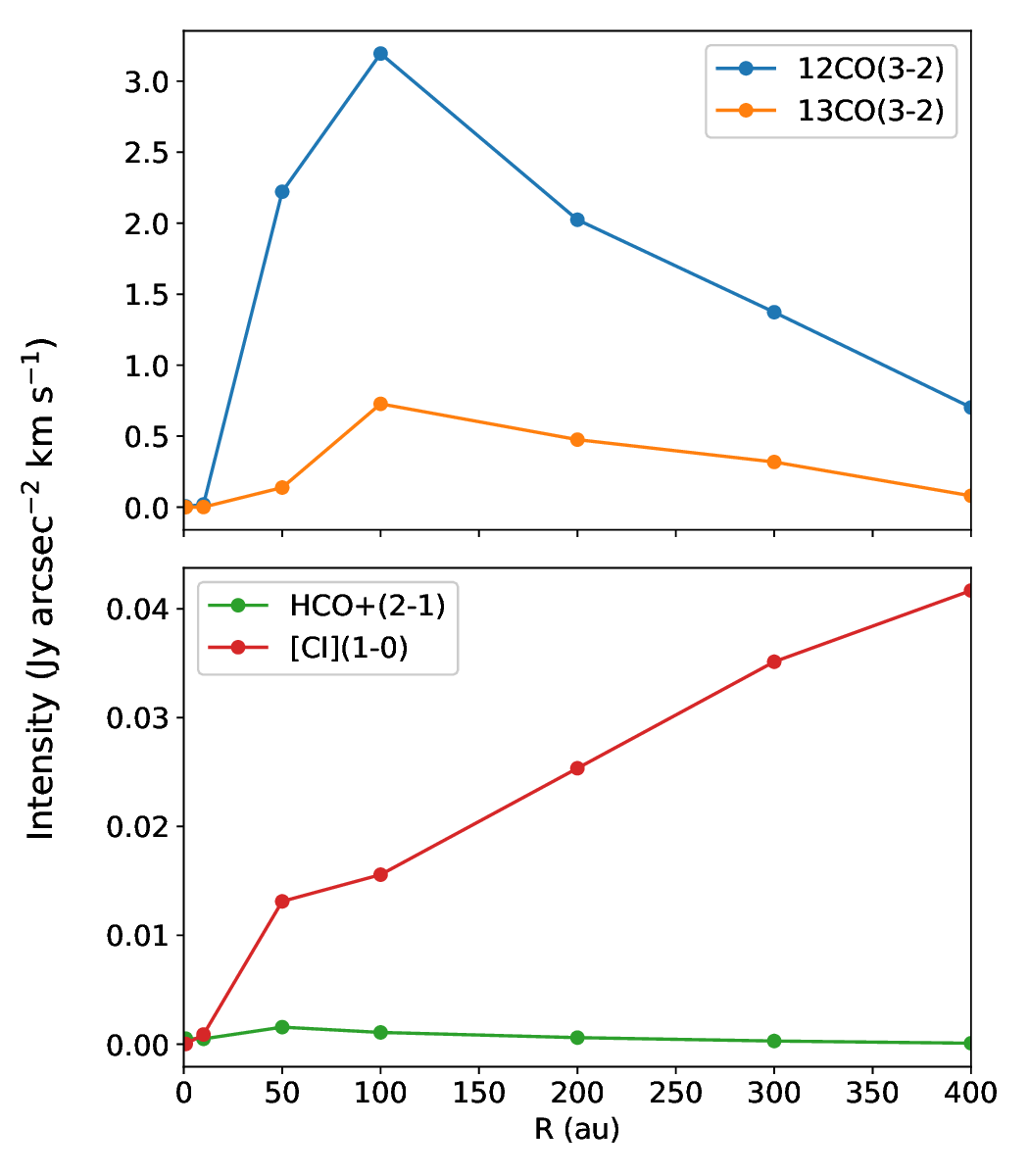}
    \caption{Radial profiles of the line intensities in Jy\,arcsec$^{-2}$\,km\,s$^{-1}$ assuming the distance to HD~121617 ($d=117.9$\,pc). The profiles show the azimuthally averaged emission as a function of radius for $^{12}$CO($J\!=\!3$--2), $^{13}$CO($J\!=\!3$--2), HCO$^{+}$($J\!=\!2$--1), and [C\,\textsc{i}] ($^{3}P_{1}$--$^{3}P_{0}$).}
    \label{fig:low_CR_line}
\end{figure}

Figure~\ref{fig:low_CR_line} shows the radial intensity profiles assuming the distance to HD~121617 ($d=117.9$\,pc). We convert energy flux to integrated line flux density and then compute the surface brightness per sky area of the annulus.
Our model reproduces the observed order of magnitude of the $^{12}$CO($J\!=\!3$--2) and $^{13}$CO($J\!=\!3$--2) radial intensities in ARKS measurements \citep{MacManamon_2026} even though our models are not tuned to the particular system. The model, however, predicts stronger \CI{} ($^{3}P_{1}$--$^{3}P_{0}$) emission at large radii, resulting in a more extended radial distribution than inferred from previous observations \citep{Cataldi_2023}.

\begin{figure}[h]
    \centering
    \includegraphics[width=1.\linewidth]{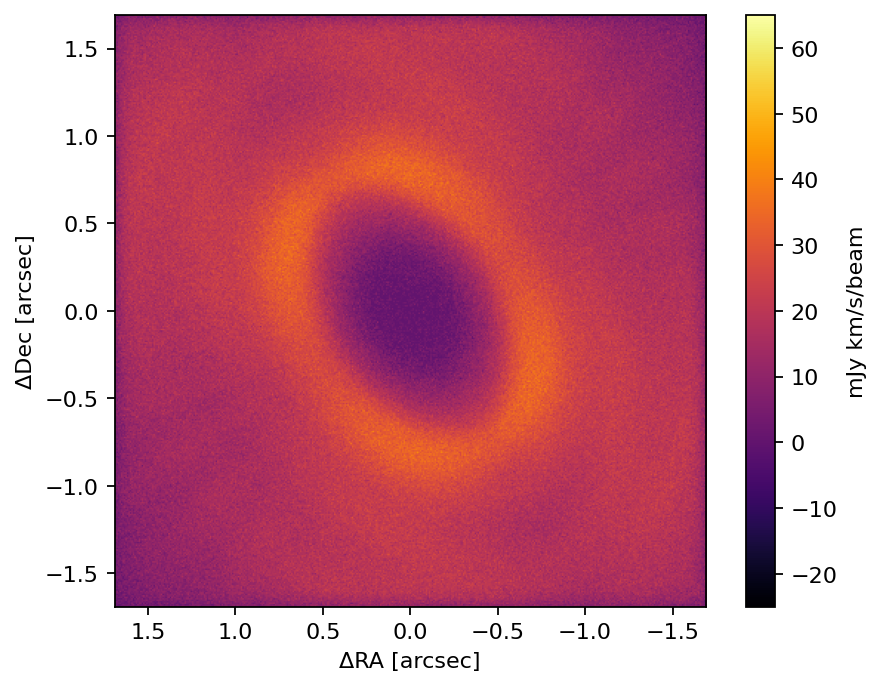}
    \caption{\tCO{} integrated-intensity map for the representative weak-CR, 40 Myr model. The image is rendered using the viewing geometry, distance, and beam size adopted from HD~121617 in order to provide an illustrative observational fiducial \citep{Brennan_2026}.}
    \label{fig:synthetic_image}
\end{figure}

Figure~\ref{fig:synthetic_image} shows the synthetic image map of the representative weak-CR, 40 Myr model, rendered using the viewing geometry and distance adopted from HD~121617 for illustration.
The rendered \tCO{} morphology is qualitatively similar to the observed ring-like emission, showing that the representative primordial-remnant model can produce a plausible resolved appearance under current observing conditions. At the same time, the model emission remains too extended, especially in \CI{}. In our model, \CI{} emission arises from regions beyond 300 au, whereas the observations \citep{Cataldi_2023} indicate a much more compact distribution, with \CI{} confined to radii of about 100 au.
The present setup should be regarded as illustrative rather than as a satisfactory source-specific description. 

This mismatch in the radial extent is important: while the disk-integrated \CI{} mass can remain low enough to satisfy current constraints, the spatial distribution of \CI{} provides a more stringent test of the model. The overly extended \CI{} emission may partly reflect our one-dimensional treatment of the external radiation field, which neglects radial ISRF irradiation from the outer disk edge; we discuss its possible effects on the outer structure in Sect.~\ref{sec:irradiation_limitation}.

\section{Discussion}
\label{sec:discussions}
In this section, we discuss the implications of our results in the context of observations of gas-rich debris disks. We first compare the model predictions with the available observational constraints and assess the plausibility of a primordial-origin scenario (Sect.~\ref{sec:plausibility}). We then examine how the main conclusions depend on key physical and chemical ingredients, in particular, the cosmic-ray ionisation rate, which can substantially affect disk chemistry (Sect.~\ref{sec:cr_rate}). 

\subsection{Plausibility of the primordial-remnant scenario}
\label{sec:plausibility}

Recent observations have revealed low \CI/CO ratios in several gas-rich debris disks, which are challenging to reconcile with current secondary-origin models (e.g., HD~110058, HD~131488, and HD~32297; \citealt{Brennan_2024}; \citealt{Cataldi_2020, Cataldi_2023}). In contrast, our primordial-origin models reproduce low disk-integrated \CI/CO ratios over a substantial part of the explored parameter space. Across our dust-poor models, the disk-integrated ratio spans $M_{\mathrm{C\,I}}/M_{\mathrm{CO}} \sim 10^{-4}$--$10^{-1}$, with the lowest values found in younger disks, where UV penetration is reduced and CO remains efficiently shielded. In this sense, the observed combination of strong CO emission and weak \CI{} emission is broadly consistent with a primordial-remnant interpretation.

The very low ratios predicted for the youngest models ($\sim$20 Myr) are smaller than the values inferred in systems where both CO and \CI{} have been detected. However, this does not necessarily imply a discrepancy. In our models, the CO $J=3\!-\!2$ and $^{13}$CO $J=3\!-\!2$ lines are optically thick over a wide radial range ($R \simeq 10$--300 au), and the CO mass residing in optically thin regions accounts for less than 1\% of the total CO mass, while the optically thin $^{13}$CO component still traces less than 10\% of the total $^{13}$CO mass. Observational estimates of the CO mass based on optically thin assumption may therefore underestimate the total CO reservoir, which would bias the inferred \CI/CO ratio high. This possibility is also supported by the recent ARKS analysis of HD~121617, in which the $^{13}$CO emission is inferred to be optically thick and the CO mass estimate is correspondingly model dependent \citep{Brennan_2026}. Conversely, the non-detection of \CI{} in some systems remains compatible with the very low ratios predicted by our younger primordial-remnant models.

An independent argument for a large primordial (or hybrid) H$_2$ reservoir comes from recent spatially resolved observations of HD~121617. Hydrodynamical modelling of the millimetre continuum arc suggests that, if the observed asymmetry is driven by gas--dust coupling, the total gas mass must be substantial, $M_{\rm gas} \sim 2.5$--250~$M_\oplus$ \citep{Weber_2026}. Compared with constraints on the CO mass of the same system ($M_{\rm CO} \sim 0.15\,M_\oplus$; \citealt{Brennan_2026}), this implies a very large H$_2$/CO ratio, which is naturally accommodated in primordial or hybrid scenarios. These dynamical estimates are broadly compatible with the gas masses in our primordial-origin models, although they do not uniquely favour that interpretation.

At the same time, the inferred mean molecular weight in HD~121617 ($\mu = 12.6^{+1.3}_{-1.1}$; \citealt{Brennan_2026}) suggests a composition that may not be dominated by H$_2$, highlighting a potential tension between the dynamical and chemical inferences. In \citet{Brennan_2026}, the mean molecular weight was estimated by fitting the data cube with the same hydrostatic disk model as our Eq.~\ref{eq:scale_height}. Although this estimate is not directly comparable to ours, we can derive an effective scale height from our CO density profile. 

A Gaussian fit to the $n_{^{13}{\rm CO}}(z)$ profile yields $H = 7.7$--9.6 au, corresponding to an effective $\mu = 2.3$--4.4 at $R=100$ au. This value is higher than expected for a fully H$_2$-dominated disk, but remains well below the value inferred by \citet{Brennan_2026}. The key point is that, even if molecular hydrogen dominates the total gas mass, the CO emission can arise from a vertically confined layer because of photodissociation, and the inferred mean molecular weight is therefore highly model dependent. In the framework discussed by \citet{Rosotti_2025}, the emitting height of an optically thick CO line depends on the thermal and column-density structure of the CO-rich layer and does not need to provide a one-to-one tracer of the hydrostatic scale height of the bulk gas. In this sense, the scale height inferred from CO emission is more likely to reflect the effective height of the CO-emitting layer than that of the full gas distribution. 

Recent near-infrared absorption spectroscopy provides an additional and more direct constraint on the molecular hydrogen content of CO-rich debris disks. \citet{Smith_2026} searched for the molecular hydrogen and CO rovibrational absorption lines toward the nearly edge-on disks HD~110058 and HD~131488, deriving lower limits on the ratio of the column densities $N_{\rm CO}/N_{\rm H_2}>1.35\times10^{-3}$ and $>3.09\times10^{-5}$, respectively. These measurements are particularly relevant because a purely H$_2$-dominated primordial disk with an approximately interstellar CO/H$_2$ abundance ratio would predict a much smaller value. Our fiducial dust-poor models with ${\rm DTG}=10^{-4}$ indeed remain H$_2$-dominated and therefore do not generally reach the high CO/H$_2$ ratio inferred for HD~110058. This is a constraint on the H$_2$-rich version of the primordial-remnant model rather than on the broader possibility of residual primordial gas. If the surviving hydrogen reservoir is partly converted from H$_2$ to atomic H in an extremely dust-depleted disk, a high CO/H$_2$ ratio can be obtained without eliminating primordial hydrogen altogether.

To assess whether the high CO/H$_2$ ratios inferred from absorption spectroscopy can be reached within the primordial-remnant framework, we performed one additional diagnostic calculation based on the fiducial externally irradiated model, changing only the dust-to-gas mass ratio to an extreme value of ${\rm DTG}=10^{-8}$. 
This run is not part of the main parameter grid, but is intended to isolate the chemical response to almost complete small-dust depletion. In this extreme limit, the molecular hydrogen reservoir is strongly converted into atomic hydrogen, while a non-negligible amount of CO can survive. 
The resulting gas is therefore no longer H$_2$-dominated, even though hydrogen remains abundant, and the CO/H$_2$ ratio can increase to $\sim6\times10^{-3}$, exceeding the lower limit for HD 110058.

This result does not imply that all CO-rich debris disks require such extreme dust depletion. 
Rather, it demonstrates that a high CO/H$_2$ ratio alone does not necessarily rule out residual primordial hydrogen, because the surviving hydrogen reservoir may be predominantly atomic rather than molecular.
 
This point suggests that future constraints on atomic hydrogen would provide a useful complement to H$_2$ absorption searches. In a secondary-origin scenario, both H$_2$ and the total hydrogen reservoir are expected to be relatively depleted, except for hydrogen produced by photodissociation of volatile-bearing exocometary material. In contrast, an extremely dust-depleted primordial remnant could show a high CO/H$_2$ ratio while still retaining a substantial reservoir of hydrogen. Observations that constrain atomic hydrogen, together with CO, \CI, and H$_2$, may therefore help distinguish genuinely secondary gas from a highly dust-depleted primordial or hybrid remnant.
Direct constraints on H and H$_2$ would provide the most informative test of whether a substantial primordial hydrogen reservoir remains.

\subsection{Implication for the cosmic-ray ionisation rate}
\label{sec:cr_rate}
The results in Sect.~\ref{sec:cr_ionisation} indicate that the current \HCOp{} upper limits favour strongly suppressed cosmic-ray ionisation within our externally irradiated primordial-remnant models.
This should be interpreted as a constraint on the ionisation environment within the adopted model framework, rather than as a unique requirement of the primordial-origin scenario itself.
The important point is that \HCOp{} responds much more strongly to $\zeta_{\rm CR}$ than \CO{} or \CI{}.
Therefore, the absence of detectable \HCOp{} does not necessarily contradict the presence of a substantial CO reservoir; instead, it suggests that the molecular-ion chemistry is likely inefficient in the CO-emitting region.
The physical origin of such a low ionisation rate remains uncertain.
In protoplanetary disks, reduced cosmic-ray ionisation can arise if stellar winds or magnetic fields exclude low-energy cosmic rays \citep{Cleeves_2013,Cleeves_2015,Fujii_2022}.

It is also important to note that the gas column in the wind is far too small to attenuate cosmic rays.
For a roughly spherical or wide-angle outflow, the column density of the wind $N_{\rm w}$ can be given as:
\begin{equation}
\begin{split}
N_{\rm w}
\sim&
\frac{\dot{M}_{\rm w}}{4\pi \mu m_{\rm H} v_{\rm w} R}\\
\simeq&
1.4\times10^{15}
\left(\frac{\dot{M}_{\rm w}}{10^{-12}\,\Msun\,{\rm yr}^{-1}}\right)
\left(\frac{v_{\rm w}}{10\,{\rm km\,s}^{-1}}\right)^{-1}\\
&\times\left(\frac{R}{100\,{\rm au}}\right)^{-1}
\left(\frac{\mu}{1.4}\right)^{-1}
{\rm cm}^{-2},
\end{split}
\end{equation}
where $\dot{M}_{\rm w}$ and $v_{\rm w}$ represent the mass loss rate and the wind velocity.
This is many orders of magnitude smaller than the characteristic surface densities required for substantial cosmic-ray attenuation, which are of order $\sim10^{25}-10^{26}\,{\rm cm^{-2}}$ depending on the particle energy and transport regime \citep{Padovani_2018}.
Thus, the wind column itself cannot be responsible for the low $\zeta_{\rm CR}$ inferred from the \HCOp{} constraints.
If the cosmic-ray ionisation rate is reduced to $\zeta_{\rm CR}\sim10^{-19}\,{\rm s}^{-1}$, the suppression would have to arise from a process other than simple gas-column attenuation, such as modulation by magnetised stellar or disk winds \citep{Cleeves_2013,Cleeves_2015}.

For gas-rich debris disks around A-type stars, however, the level of stellar magnetic activity and the efficiency of cosmic-ray exclusion are much less clear.
The low value inferred here, $\zeta_{\rm CR}\sim10^{-19}\,\mathrm{s}^{-1}$, should therefore be regarded as a testable requirement of the present primordial-remnant interpretation, not as an established property of these systems.

Observationally, deeper searches for \HCOp{} and other molecular ions, together with spatially resolved CO and \CI{} maps, would test the weak CR ionisation chemistry and radial gas structure predicted by the present primordial-origin models.

\section{Model limitations}
\label{sec:limitation}
Our models include several simplifying assumptions, which we discuss in this section. These limitations affect not only the technical setup of the calculations but also how confidently the predicted \CO{}, \CI{}, and \HCOp{} masses and spatial distributions can be compared with observations. 

\subsection{Thermal structure}
\label{sec:thermal_limitation}
Our thermochemical post-processing separates the prescribed thermal structure from the radiation field used for photochemistry. The gas temperature is taken from the O25 disk-evolution model, which includes stellar irradiation and viscous heating, whereas the incident radiation field in \textsc{Cloudy} is varied to explore limiting photochemical environments. This treatment is therefore not fully thermochemically self-consistent.

As a diagnostic, we also performed calculations using \textsc{Cloudy}'s thermal balance. In the internal-irradiation case, which includes stellar radiation, \textsc{Cloudy} predicts a midplane temperature of $T_{\rm mid}\simeq40\,{\rm K}$ at $R\sim100$ au. In the external-only case, the predicted temperature falls to approximately $10\,{\rm K}$ at the same radius. This low temperature leads to efficient CO freeze-out because the external-only calculation omits stellar radiative heating. We therefore regard the O25 temperature as a more appropriate thermal background for the present bracketing calculations.

Our sensitivity tests suggest that the main qualitative results are not determined solely by this temperature prescription. Changing the prescribed temperature by 30\% modifies the disk-integrated masses of CO, \CI{}, and \HCOp{} by at most approximately 20\%. In the internal-irradiation case, using the prescribed O25 temperature instead of the \textsc{Cloudy} thermal-balance solution changes the midplane CO gas density typically by a factor of approximately two. Nevertheless, the absolute molecular masses, \CI{}/CO ratios, and line intensities remain uncertain at approximately the order-unity level.

In addition, the adopted disk structure is vertically isothermal at each radius. In the externally irradiated models, the surface layers would in reality be warmer than the midplane, which could enhance the excitation and emergent line intensities and, if the vertical density structure were recomputed in hydrostatic equilibrium, increase the gas scale height and make the atomic-H layer more vertically extended. A self-consistent calculation including both stellar and external irradiation is therefore required before these models can be used for precision gas-mass estimates.

\subsection{Irradiation treatment}
\label{sec:irradiation_limitation}
Our treatment of irradiation involves two distinct approximations. The first concerns the geometry of the incident radiation field, whereas the second concerns the conversion of the stellar FUV continuum into the CO photodissociation rate. These approximations affect different aspects of the model predictions.

We consider only two limiting irradiation geometries: pure external irradiation by the interstellar radiation field and pure internal irradiation by the host star. Real disks are likely to lie between these two extremes, with the inner disk more strongly affected by stellar irradiation and the outer disk more strongly influenced by external irradiation. This simplification is expected to affect the radial distribution of chemical species. In the internally irradiated models, CO is efficiently photodissociated at small radii, whereas in the externally irradiated models CO can remain abundant in the inner disk. By contrast, observed CO-rich debris disks often show CO emission concentrated around $R\sim100$ au, suggesting that a radially varying mixture of internal and external irradiation would provide a more realistic description of the emitting region.

The internally irradiated models have an additional uncertainty associated with the treatment of CO photodissociation. In the \textsc{Cloudy} version used here, the CO photodissociation rate is parameterised using an unshielded rate calibrated for an interstellar Habing field and is scaled with the broadband FUV intensity, together with dust attenuation and CO self-shielding \citep{2012_UMIST,Ferland_2017,Chatzikos_2023}. This prescription does not explicitly convolve the incident stellar spectrum with the discrete CO absorption bands between 911.8 and 1117.8~\AA. Because the ratio of the CO-dissociating photon flux to the broadband FUV flux depends on spectral shape, applying this ISRF-calibrated scaling to an A-star spectrum can introduce a large systematic uncertainty.

To estimate the magnitude of this uncertainty, we performed an approximate spectral-rescaling experiment for the internal-only model with DTG$=10^{-4}$, the weak cosmic-ray ionisation rate, and the prescribed gas temperature. Rescaling the stellar FUV continuum according to its relative flux in the CO-dissociating band lowers the disk-integrated \HCOp{} mass by more than two orders of magnitude. In the calculation using the standard prescription, efficient CO photodissociation maintains C- and C$^+$-rich irradiated layers, where \HCOp{} is formed mainly through pathways involving \CHtp{} and CO$^+$. The rescaling instead leaves most carbon in CO and suppresses these ion--molecule formation pathways. This experiment should not be interpreted as a precise correction to the internal-only model. The rescaling modifies all photoprocesses within the selected FUV interval rather than the CO photodissociation rate alone, and it does not treat the individual CO absorption lines or their wavelength-dependent shielding. Nevertheless, it demonstrates that the large \HCOp{} abundance obtained in the uncorrected internal-only calculation is highly sensitive to the CO photodissociation prescription and is not a robust consequence of stellar irradiation.

The external-irradiation models have a separate geometrical limitation: the ISRF is applied only along the vertical direction at each radius. In a real disk, ISRF photons can also penetrate radially from the outer disk edge. As an order-of-magnitude estimate, we integrated the neutral-carbon density of the representative weak-CR external-irradiation model from the outer disk edge inward. Along the midplane, this gives a radial \CI{} column of only $N_{\rm C\,I,rad}\lesssim3\times10^{16}\,\mathrm{cm^{-2}}$. This additional radial irradiation would preferentially affect the low-column-density outer disk, where the radial shielding column is small, and could therefore move the outer CO photodissociation front inward. The effect on the \CI{} emission is not necessarily monotonic: enhanced CO photodissociation can produce neutral carbon in the transition layer, but the most exposed gas may be further photoionised to C$^+$. A neutral-carbon column density below $N_{\rm C\,I}\sim10^{17}\,\mathrm{cm^{-2}}$ is insufficient to provide effective shielding against C photoionisation \citep{Heays_2017}. Thus, including radial ISRF irradiation could either enhance a narrow outer \CI{} rim or, if the exposed carbon becomes predominantly ionised, suppress the extended neutral-carbon emission and make the \CI{} distribution more compact.  Because our representative model overpredicts the radial extent of \CI{}, this missing radial irradiation may act in the direction needed to reduce the outer \CI{} emission, but a genuine two-dimensional treatment of the ISRF, including CO self-shielding and C/C$^+$/CO transitions, is required to determine the sign and magnitude of the effect.

Thus, the internal-only \HCOp{} mass and the external-model radial profiles are uncertain for different reasons. Improving the former requires an SED-dependent, preferably line-by-line treatment of CO photodissociation, whereas improving the latter requires a multidimensional calculation including both stellar and interstellar irradiation.

\subsection{1D disk evolution model}
\label{sec:lim_disk_evolution}
Our chemistry is computed in post-processing and is not coupled self-consistently to the long-term disk evolution. The surface-density evolution model provides the background density and temperature structure, while the chemical abundances are calculated afterwards. Because the thermal and chemical structure can in turn influence the viscosity, ionisation balance, and gas dispersal history, a fully coupled chemo-dynamical treatment could modify both the absolute gas masses and the disk lifetime. This limitation is especially important for the predicted spatial extent of the gas. One of the remaining discrepancies in our models is not the disk-integrated \CI{}/CO ratio, but the fact that the synthetic \CI{} emission is more radially extended than suggested by current observations.

The underlying disk-evolution model depends on the adopted turbulent-transport prescription, magnetic angular-momentum transport, and the initial and outer-boundary conditions. In addition, the disk-evolution model assumes that small grains are depleted throughout the disk, including its outermost regions. A radially varying degree of small-grain depletion could modify the thermal and dynamical evolution of the outer gas and hence the resulting \CI{} distribution. 

A more realistic evolutionary history could include an earlier phase in which small grains were still abundant and FUV-driven photoevaporation truncated part of the outer disk, followed by small-grain depletion and the onset of the long-lived FUV-weak phase. Alternatively, the disk could have been dynamically truncated by an early stellar encounter in its birth environment \citep{Winter_2018}. These possibilities would modify the outer boundary condition inherited by the long-lived primordial remnant rather than invalidate the small-grain-depleted framework itself. Future evolutionary calculations should test whether such pre-processing can produce a more compact \CI{} distribution while retaining the substantial CO reservoir required by the observations.

Overall, these limitations do not remove the qualitative result that a primordial-remnant disk can retain substantial CO while maintaining low \CI{} and weak \HCOp{} under favourable ionisation conditions. They do, however, limit the precision with which the present models can predict absolute gas masses, line intensities, and source-specific radial distributions. A more self-consistent treatment of irradiation, thermal balance, chemistry, and gas dispersal will therefore be needed to test the primordial-remnant interpretation in detail.

\section{Conclusions}
\label{sec:conclusions}

We investigated whether a long-lived primordial gas disk can account for the key chemical constraints of CO-rich debris disks around intermediate-mass stars at ages of tens of Myr. Using the 1D evolution models for disks around a $2\,M_\odot$ star \citep{Ooyama_2025}, we post-processed representative disk structures at $20$--$40$ Myr with the thermochemical code \textsc{Cloudy}. We derived the chemical structures in two cases: (i) externally irradiated by an interstellar radiation field and (ii) internally irradiated by the host star. We varied the dust-to-gas mass ratio, dust size distribution, and the cosmic-ray ionisation rate in order to jointly constrain CO, \CI, and \HCOp.

Our main findings are as follows.

\begin{enumerate}
\item
Dust-poor primordial-remnant disks can retain substantial CO at tens of Myr.
In the low-DTG models, CO survives in shielded outer-disk regions and can remain optically thick around $\sim100$ au (Sect.~\ref{sec:fid}; Fig.~\ref{fig:fid_column}).
This supports the chemical viability of CO-rich primordial-remnant disks at ages of tens of Myr, although the absolute CO mass remains tied to the adopted disk-evolution model and radiative-transfer assumptions.

\item 
In the externally irradiated disks, the \HCOp{} mass and $J=2\rightarrow1$ luminosity are controlled mainly by midplane ion--molecule chemistry initiated by cosmic-ray ionisation. A direct comparison with current line-luminosity upper limits disfavours the standard cosmic-ray ionisation rate in the low-DTG models and favours suppressed values of order $\zeta_{\rm CR}\sim10^{-19}\,\mathrm{s}^{-1}$ within our adopted framework (Sect.~\ref{sec:cr_ionisation}; Fig.~\ref{fig:mhcop_zeta}), although source-specific modelling is still required for individual systems.

The internal-irradiation limit produces \HCOp{} efficiently in exposed upper layers (Sect.~\ref{sec:internal}; Fig.~\ref{fig:fid_internal_density}), showing that the \HCOp{} constraint is sensitive to the internal irradiation. However, an approximate spectral-rescaling experiment reduces the disk-integrated \HCOp{} mass by more than two orders of magnitude (Sect.~\ref{sec:irradiation_limitation}). This sensitivity suggests that the default \texttt{Cloudy} CO photodissociation prescription may significantly overestimate the \HCOp{} mass when applied to the spectra of early A-type stars. The uncorrected \HCOp{} mass should therefore be regarded as a high-end estimate rather than a robust prediction. A self-consistent multidimensional treatment is required to refine this constraint.

\item 
The comparison between modelled and observed \CI{}/\CO{} ratios must account for the fact that CO emission can be optically thick in CO-rich debris disks.
If the observationally inferred CO mass is a lower limit, the inferred \CI{}/\CO{} ratio is biased high, and the true ratio may be lower than the apparent value.
With this caveat, our models can reproduce low disk-integrated \CI{}/\CO{} ratios, with typical values of $M_{\CI}/M_{\CO}\lesssim0.1$ and values down to $\sim10^{-3}$ in the youngest, most strongly shielded cases.

However, the agreement is limited to the disk-integrated ratio.
The representative model predicts \CI{} emission that is more radially extended than suggested by current observations (Fig.~\ref{fig:low_CR_line}).
Thus, the main remaining question is whether the adopted primordial-origin disk structure can reproduce the observed \CI{} distribution.

\end{enumerate}

Taken together, our results support the primordial-origin scenario as a viable explanation for CO-rich debris disks: in dust-poor disks with weak CR ionisation, it can simultaneously account for substantial CO, low \CI/\CO{} total mass ratios, and weak \HCOp{} emission.

The extended \CI-emitting region in our representative model should not yet be interpreted as a robust failure of this scenario. Its radial extent may depend both on the omission of radially incident ISRF photons and on the outer gas structure inherited from the 1D evolution model. The latter remains uncertain because of the adopted transport prescription, magnetic angular-momentum transport, boundary conditions, and the assumed radial extent of small-grain depletion. Testing the robustness of the predicted \CI{} distribution will require multidimensional irradiation calculations and a broader exploration of plausible outer-disk structures.

Distinguishing among primordial, secondary, and hybrid interpretations will require joint constraints on the total hydrogen reservoir and the partitioning of carbon between CO and \CI, together with molecular-ion tracers such as \HCOp{}, interpreted using detailed chemical models.

\begin{acknowledgements}
HM has been supported by JSPS Overseas Research Fellowship. This work was supported by JSPS KAKENHI Grant Number 25K17432 (HM) and 19KK0353 (TH). We thank Yuri Aikawa, Aoife Brennan, Gianni Cataldi, and Kazunari Iwasaki for insightful discussions. This work was also supported by JST SPRING (WO: JPMJSP2110). W.O. is supported by the Graduate School of Science, Kyoto University, under the Ginpu Fund. R.N. acknowledges support from the European Union (ERC Starting Grant DiscEvol, project number 101039651) and from Fondazione Cariplo, grant No. 2022-1217.
RK acknowledges financial support via the Heisenberg Research Grant funded by the Deutsche Forschungsgemeinschaft (DFG, German Research Foundation) under grant no.~KU 2849/9, project no.~445783058. TH appreciates the financial support from the Kyoto University Foundation.
\end{acknowledgements}

\bibliographystyle{aa}
\bibliography{references}

\clearpage

\onecolumn
\begin{appendix}
\nolinenumbers    

\section{Chemical reaction rates}
We summarise the chemical reaction rates in midplane of the fiducial case.
\begin{figure*}[h]
    \centering
    \includegraphics[width=1.\linewidth]{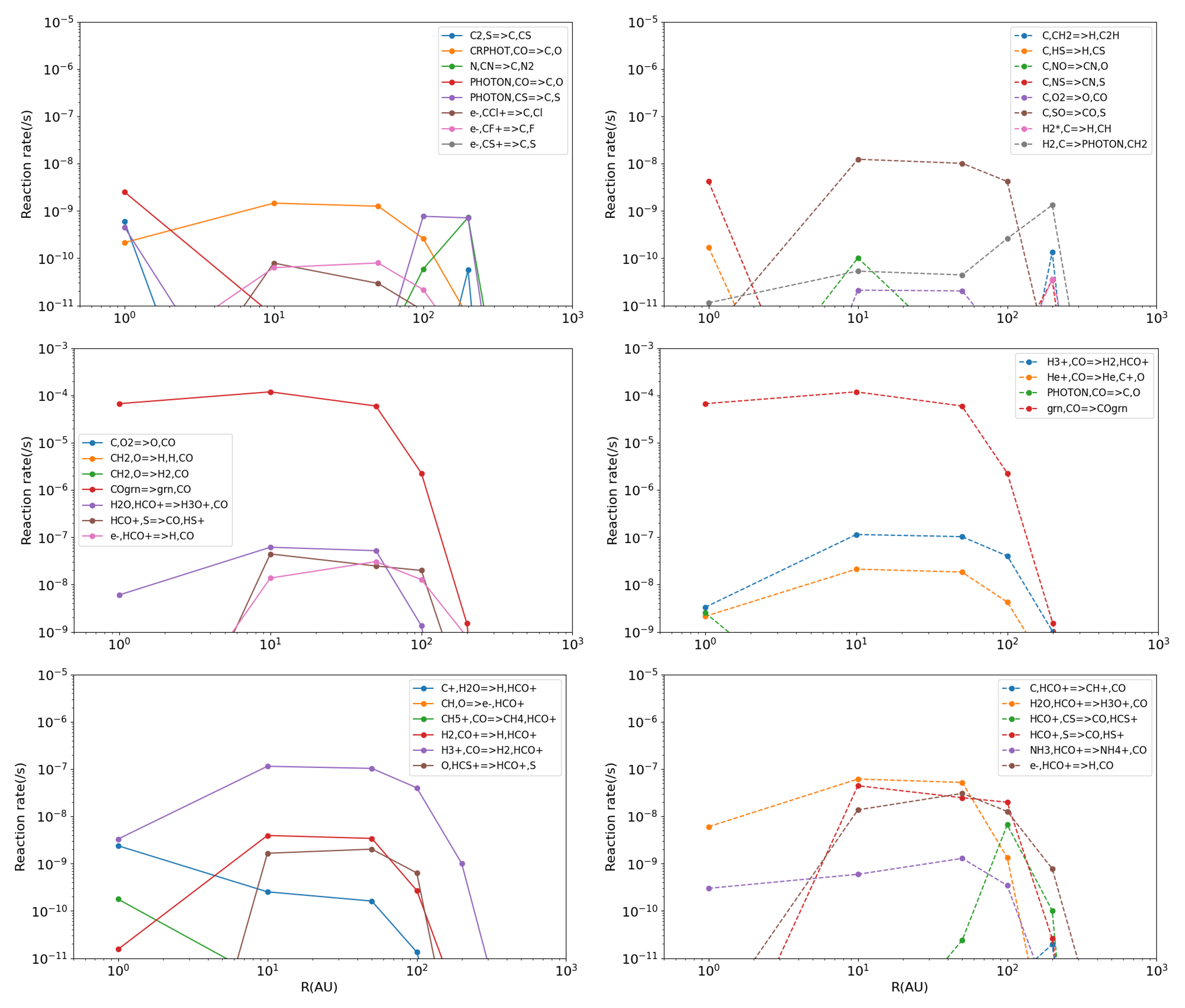}
    \caption{Reaction rates in the midplane of the fiducial case. \CI{} related reactions (top), \CO{} reactions (middle), and \HCOp{} reactions (bottom) are shown. The left column shows the formation reaction (solid lines), while the right column shows the destruction reaction (dashed lines). }
    \label{fig:fid_reaction}
\end{figure*}

\section{Disk-integrated species masses}
\label{app:species_masses}

Table~\ref{tab:species_masses} summarises the disk-integrated species masses for
the externally irradiated models used in this work. 

\begin{table}[p]
\caption{Disk-integrated species masses for the external-irradiation models.}
\label{tab:species_masses}
\centering
\begin{tabular}{ccccccccc}
\hline\hline
$t$ & DTG & $\zeta_{\rm CR}$ & H$_2$ & H & \CO{} & $^{13}$\CO{} & \CI{} & \HCOp{} \\
$[\Myr]$ & & $[\mathrm{s}^{-1}]$ & \multicolumn{6}{c}{Mass [$\Mearth$]} \\
\hline
20 & $10^{-2}$ & $10^{-17}$ & $616$ & $0.160$ & $0.956$ & $0.0330$ & $2.29\times10^{-3}$ & $4.85\times10^{-7}$ \\
20 & $10^{-2}$ & $10^{-18}$ & $620$ & $0.151$ & $0.649$ & $0.0224$ & $2.35\times10^{-3}$ & $4.91\times10^{-8}$ \\
20 & $10^{-2}$ & $10^{-19}$ & $614$ & $0.150$ & $0.609$ & $0.0210$ & $2.34\times10^{-3}$ & $1.68\times10^{-8}$ \\
20 & $10^{-4}$ & $10^{-17}$ & $615$ & $3.56$ & $1.67$ & $0.0576$ & $1.66\times10^{-3}$ & $7.79\times10^{-7}$ \\
20 & $10^{-4}$ & $10^{-18}$ & $616$ & $2.86$ & $1.56$ & $0.0537$ & $1.28\times10^{-3}$ & $8.04\times10^{-8}$ \\
20 & $10^{-4}$ & $10^{-19}$ & $617$ & $2.80$ & $1.54$ & $0.0532$ & $1.24\times10^{-3}$ & $2.32\times10^{-8}$ \\
30 & $10^{-2}$ & $10^{-17}$ & $214$ & $0.267$ & $0.0453$ & $1.56\times10^{-3}$ & $3.98\times10^{-3}$ & $5.39\times10^{-8}$ \\
30 & $10^{-2}$ & $10^{-18}$ & $215$ & $0.251$ & $0.0382$ & $1.32\times10^{-3}$ & $4.09\times10^{-3}$ & $1.92\times10^{-8}$ \\
30 & $10^{-2}$ & $10^{-19}$ & $215$ & $0.249$ & $0.0380$ & $1.31\times10^{-3}$ & $4.09\times10^{-3}$ & $1.62\times10^{-8}$ \\
30 & $10^{-4}$ & $10^{-17}$ & $207$ & $5.88$ & $0.556$ & $0.0192$ & $3.04\times10^{-3}$ & $5.66\times10^{-7}$ \\
30 & $10^{-4}$ & $10^{-18}$ & $209$ & $4.77$ & $0.425$ & $0.0147$ & $2.42\times10^{-3}$ & $6.41\times10^{-8}$ \\
30 & $10^{-4}$ & $10^{-19}$ & $209$ & $4.66$ & $0.404$ & $0.0139$ & $2.34\times10^{-3}$ & $2.32\times10^{-8}$ \\
40 & $10^{-2}$ & $10^{-17}$ & $116$ & $0.397$ & $0.0128$ & $4.41\times10^{-4}$ & $4.92\times10^{-3}$ & $2.8\times10^{-8}$ \\
40 & $10^{-2}$ & $10^{-18}$ & $116$ & $0.375$ & $0.0119$ & $4.09\times10^{-4}$ & $5.38\times10^{-3}$ & $1.68\times10^{-8}$ \\
40 & $10^{-2}$ & $10^{-19}$ & $116$ & $0.373$ & $0.0118$ & $4.06\times10^{-4}$ & $5.45\times10^{-3}$ & $1.57\times10^{-8}$ \\
40 & $10^{-4}$ & $10^{-17}$ & $108$ & $7.96$ & $0.287$ & $9.92\times10^{-3}$ & $3.57\times10^{-3}$ & $2.3\times10^{-7}$ \\
40 & $10^{-4}$ & $10^{-18}$ & $109$ & $6.63$ & $0.199$ & $6.88\times10^{-3}$ & $3.12\times10^{-3}$ & $3.57\times10^{-8}$ \\
40 & $10^{-4}$ & $10^{-19}$ & $110$ & $6.49$ & $0.187$ & $6.46\times10^{-3}$ & $3.06\times10^{-3}$ & $2.12\times10^{-8}$ \\
40 & $10^{-8}$ & $10^{-19}$ & $3.44$ & $113$ & $0.282$ & $9.72\times10^{-3}$ & $3.15\times10^{-3}$ & $1.28\times10^{-8}$ \\
\hline
\end{tabular}
\end{table}

\end{appendix}
\end{document}